\documentclass[%
preprint,
amsmath,amssymb,
aps,
prx,
]{revtex4-1}

\usepackage{graphicx,epsfig,epstopdf}
\usepackage{amsmath}
\usepackage{color}
\usepackage{caption}
\usepackage{subfigure}
\usepackage{subfloat}
\usepackage{array}
\usepackage{tabularx}
\usepackage{multirow}
\newcolumntype{L}[1]{>{\raggedright\arraybackslash}p{#1}}
\newcolumntype{C}[1]{>{\centering\arraybackslash}p{#1}}
\newcolumntype{M}[1]{>{\centering\arraybackslash}m{#1}}

\def\e{\begin{equation}}
\def\f{\end{equation}}
\def\_#1{{\bf #1}}
\def\.{\cdot}

\begin{document}

\title{From the generalized reflection law to the realization of perfect anomalous reflectors}

\author{A.~D\'{i}az-Rubio$^{1}$, V.~S.~Asadchy$^{1}$,   
     A. Elsakka$^{1}$,
        and
   S.~A.~Tretyakov$^{1}$}
 
\affiliation{$^1$Department of Radio Science and Engineering, Aalto University, P.~O.~Box~13000, FI-00076 Aalto, Finland}

\begin{abstract}

The use of the generalized Snell's law opens wide possibilities for the manipulation of transmitted and reflected wavefronts. However, known structures designed to shape reflection wave fronts suffer from significant  parasitic reflections in undesired directions.
In this work, we explore the limitations of the existing solutions for the design of passive planar reflectors and demonstrate that strongly non-local response is required for perfect performance.  
A new paradigm for the design of perfect reflectors based on energy surface channeling is introduced. 
We realize and experimentally verify a perfect design of an anomalously reflective surface  using an array of rectangular metal patches backed by a metallic plate. This conceptually new mechanism for wavefront manipulation allows the design of thin  perfect reflectors, offering  a versatile design method applicable to other scenarios such as focusing reflectors, surface wave manipulations or metasurface holograms, extendable to other frequencies.

\end{abstract}

\maketitle

\section{Introduction}

The classical approach to the design of wave-shaping reflectors for light or microwave radiation is based on the geometrical optics. A flat mirror obviously obeys the usual reflection law: in the absence of dissipation, all reflected rays go into the specular direction (the incidence and reflection angles are equal, $\theta_{\rm i}$=$\theta_{\rm r}$) without changing the field amplitude. The distribution of reflected field intensity can be engineered by shaping the reflecting surface.
Due to the differences of the ray-propagation path lengths, the phase distribution at the reflector aperture can be tuned so that, for example, all rays converge at a point, forming a focal spot. 
Generalizing the phased-array antenna principle, the same function can be realized in a planar reflector, if the reflection phase is made non-uniform over the reflector surface.
In antenna applications, such non-uniform reflectors are called reflectarrays and are usually realized as arrays of resonant antennas \cite{encinar}. Most commonly, patch antennas are used and the reflection phase from every element is tuned either by reactive loads or by varying patch size or shape. Reflectarrays with subwavelength distances between the array elements are called high-impedance surfaces \cite{HIS,Pozar_dense_reflect_array} or  metasurfaces \cite{new_review}.

\begin{figure}
	\centering
	 \raisebox{0.5cm}{\subfigure[]{\includegraphics[width = 0.35\textheight, keepaspectratio=true]{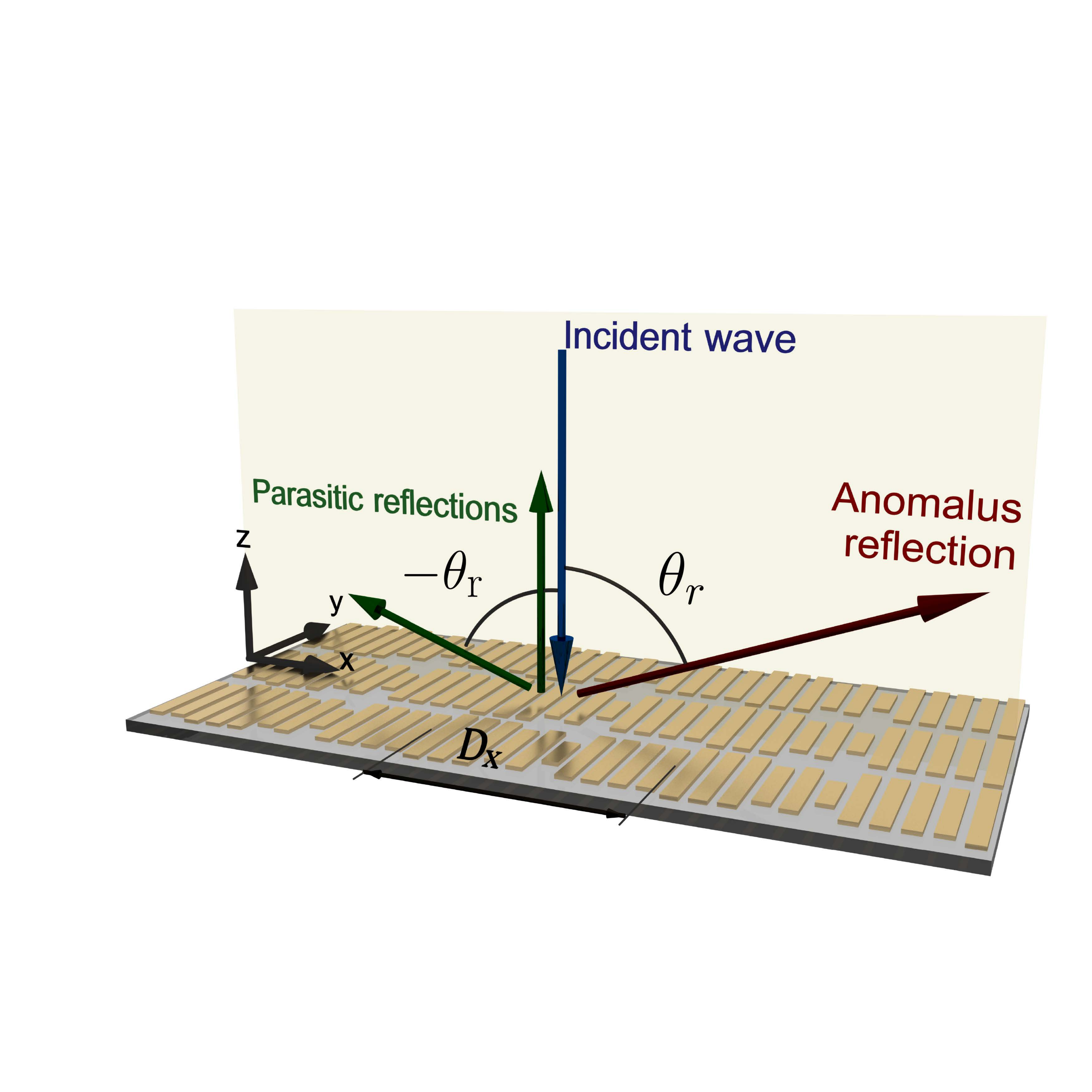}\label{fig:DiazRubioFIG1_A}}}
	\subfigure[]{\includegraphics[width = 0.3\textheight, keepaspectratio=true]{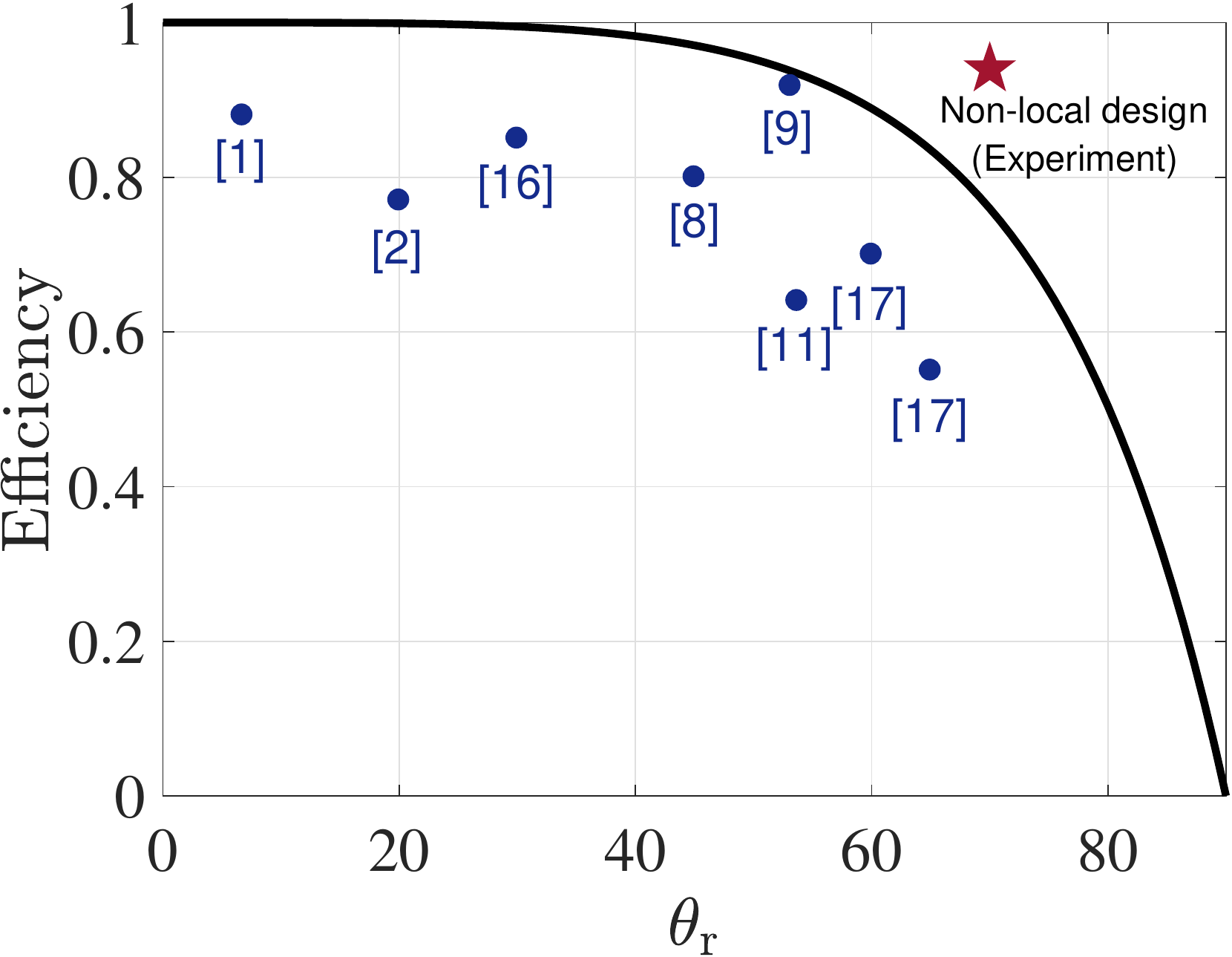}\label{fig:DiazRubioFIG1_B}}
	\caption{(a) Illustration of the performance of a reflective metasurface. Propagating waves in the system when $\theta_{\rm i}=0^\circ$ and $\theta_{\rm r}>30^\circ$ are represented in the scheme. 
(b) Comparative overview of the efficiency of anomalous-reflection metasurfaces and optical gratings. Blue dots represent previous results found in the literature; black line is the numerical estimation of the designs based on a linear $2\pi$-phase gradient calculated accordingly to  Eq.~(\ref{eq:conventional}); red star represents the results obtained in this paper.}\label{fig:DiazRubioFIG1}
\end{figure}

We consider the anomalous reflection scenario illustrated in Fig.~\ref{fig:DiazRubioFIG1_A}. According to the phased-array principle, to reflect an incident plane wave into another plane wave, breaking the usual reflection law (the reflection angle $\theta_{\rm r}\neq \theta_{\rm i}$), the reflection phase should linearly depend on the corresponding coordinate along the reflector plane. In this situation one can expect that reflections from all the points interfere constructively in a plane wave propagating in the desired direction. Recently, this simple design principle was formulated in form of the ``generalized Snell's law'' \cite{capasso}. To understand  this law, let us assume  that a plane wave is incident at a planar reflecting surface at the incidence angle $\theta_{\rm i}$ and introduce a Cartesian coordinate system with the  $x$-axis along the projection of the wavevector to the reflector plane. If the reflected plane wave is propagating at the reflection angle $\theta_{\rm r}$ and its amplitude is the same as in the incident wave, then the ratio of the tangential electric fields in the reflected wave and in the incident wave at the reflector surface is given by $\exp(j\Phi_{\rm r})=\exp[j(\sin\theta_{\rm i}-\sin\theta_{\rm r})k_1x]$. Here, $k_1=\omega\sqrt{\mu\varepsilon}$ is the wavenumber in the background isotropic medium, and we assume the time-harmonic dependency $e^{j\omega t}$. The local reflection coefficient  $
R={\left( Z_{\rm s}(x)-\eta_1\right) }/{\left(Z_{\rm s}(x)+\eta_1\right) }= \exp(j\Phi_{\rm r})$, where $\eta_1$ is the wave impedance of the incident plane wave (ratio between the tangential components of the electric and magnetic fields),  defines a periodically modulated boundary surface. The surface impedance $Z_{\rm s}(x)$, defined as the ratio between the tangential components of the total electric and magnetic fields (incident and scattered) at the surface, is purely imaginary and can be expressed as
\begin{equation} \label{eq:conventional}
	Z_{\rm s}(x)=j \frac{\eta_1}{\cos\theta_{\rm i}}\cot[{\Phi_{\rm r}(x)}/2]. 
\end{equation}
Equation (\ref{eq:conventional}) is a well-known solution that has been used in numerous works, for example, \cite{sun,bozh1,bozh2,mosall2,nanotechnology,prx4,veysi,li}. Nevertheless, the sum of the incident and one reflected plane wave is not a valid solution for the Maxwell equations with the surface impedance given by (\ref{eq:conventional}). This means that when we illuminate a metasurface characterized by surface impedance (\ref{eq:conventional}) given by the generalized reflection law with a plane wave at $\theta_{\rm i}$, in addition to the desired anomalously reflected plane wave at $\theta_{\rm r}$, more plane waves will be excited in the system in order to satisfy the power conservation and the boundary conditions, leading to parasitic reflections or energy absorption in the reflector \cite{synthesis,Alu}. 

Figure~\ref{fig:DiazRubioFIG1_B} shows the numerical estimation of the efficiency (the ratio between the power sent into the desired direction and the incident power) for a metasurface based on the generalized Snell law  modeled by the impedance boundary as in Eq.~(\ref{eq:conventional}). We  can see that in all known realizations the power efficiency is lower than the numerical prediction due to imperfections in fabrication and discretization problems. It is also worth noting that the efficiency dramatically decreases when the desired reflection angle deviates more and more from the specular reflection angle. 
At optical frequencies, traditional diffraction gratings are periodic surfaces engineered for controlling the percentage of energy reflected into each diffraction mode. In principle, for a certain incidence angle $\theta_{\rm i}$, the surface profile can be tailored in order to send the energy into some  reflection angle $\theta_{\rm r}$. These devices work efficiently for retroreflection ($\theta_{\rm i}=-\theta_{\rm r}$) or with small differences between the incidence and reflection directions ($\theta_{\rm i}\approx-\theta_{\rm r}$) \cite{Grating1,Grating2}, but the efficiency  decreases when the difference between the incidence and reflection angles increases.

\begin{figure}
	\centering
	\begin{minipage}{\linewidth}
		\subfigure[]{\includegraphics[width = 0.22\textheight, keepaspectratio=true]{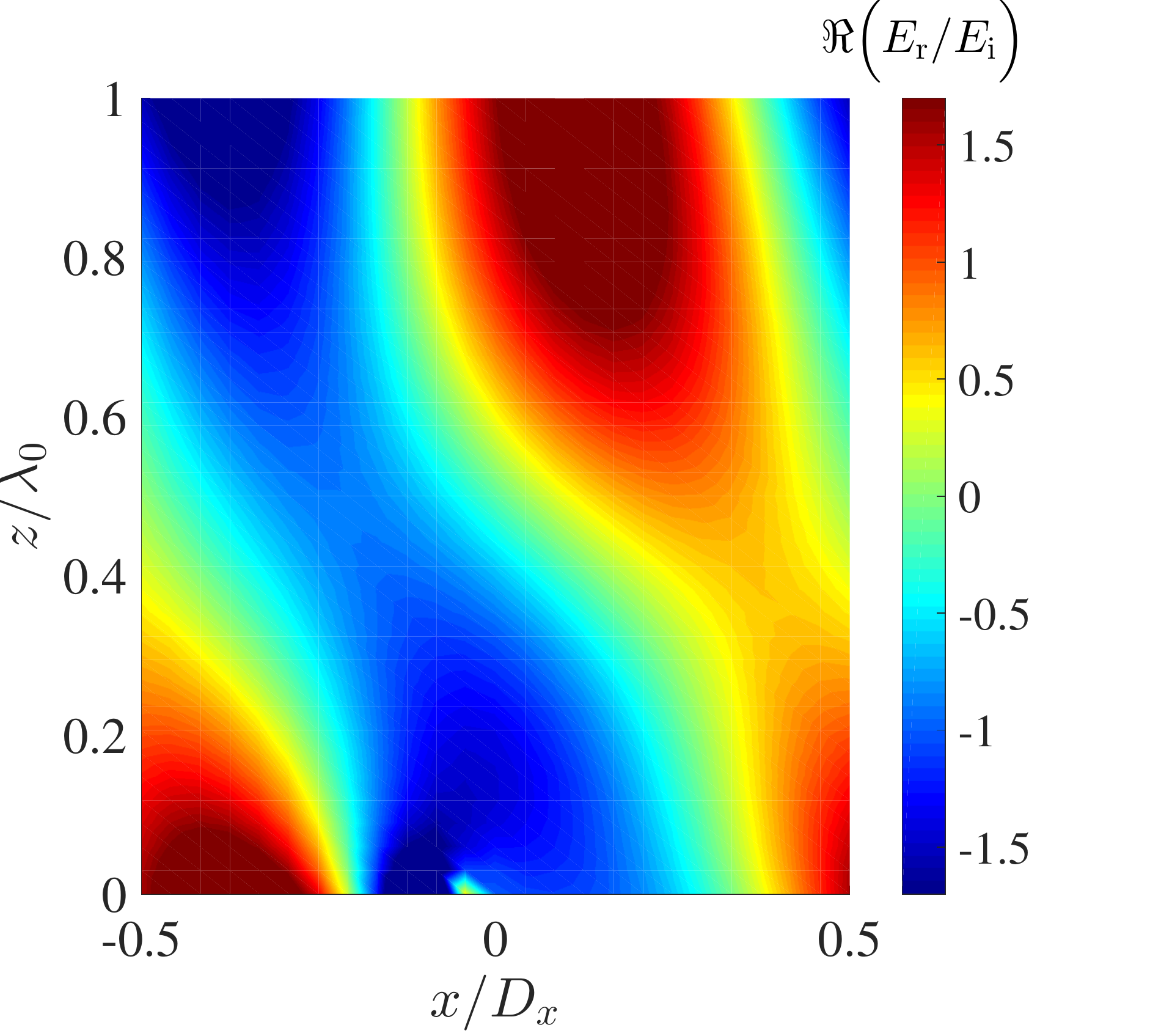}\label{fig:DiazRubioFIG2a}}
		\subfigure[]{\includegraphics[width = 0.22\textheight, keepaspectratio=true]{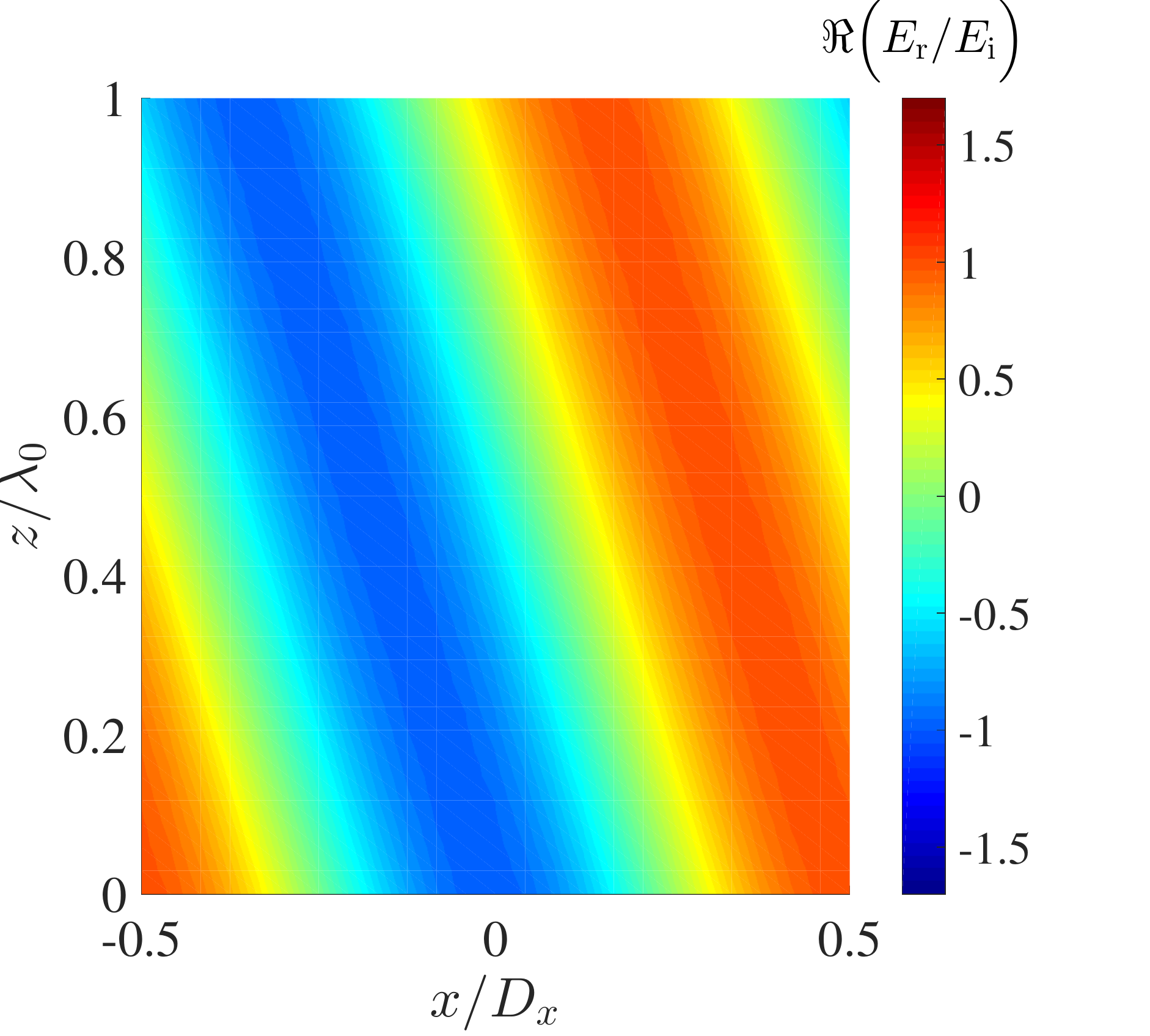}\label{fig:DiazRubioFIG2b}}
		\subfigure[]{\includegraphics[width = 0.22\textheight, keepaspectratio=true]{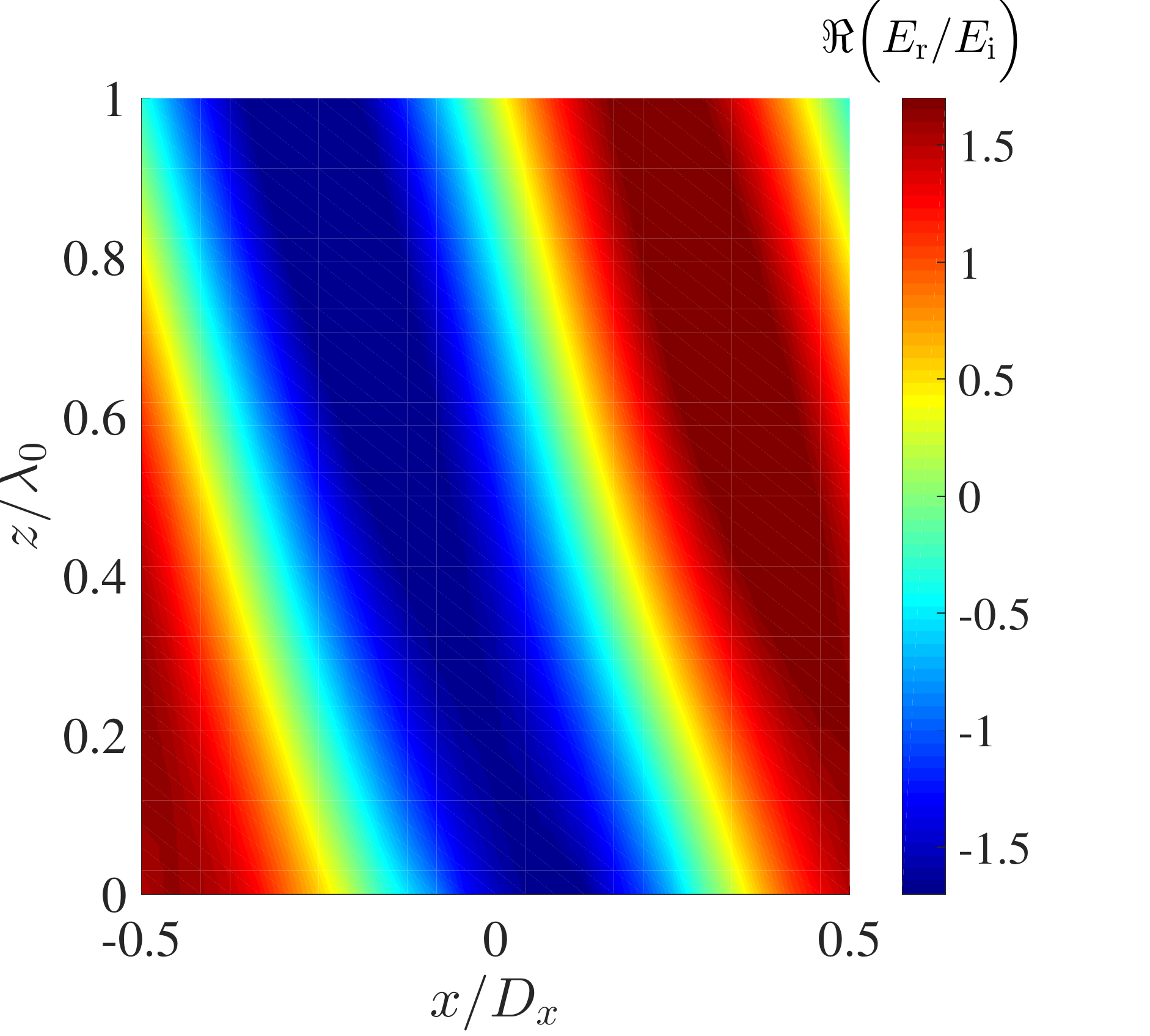}\label{fig:DiazRubioFIG2c}}
	\end{minipage}
	\begin{minipage}{\linewidth}
		\subfigure[]{\includegraphics[width = 0.22\textheight, keepaspectratio=true]{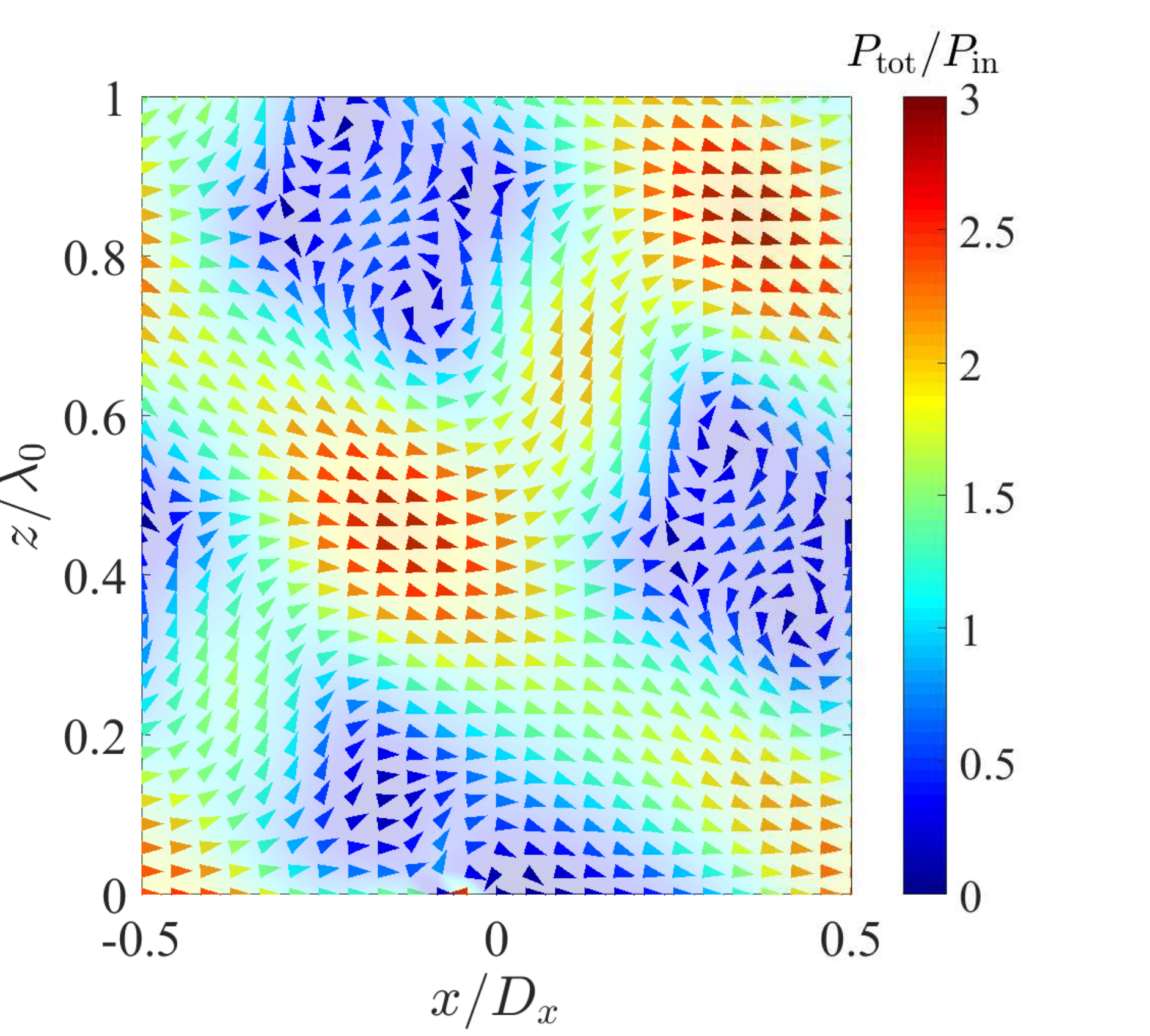}\label{fig:DiazRubioFIG2d}}
		\subfigure[]{\includegraphics[width = 0.22\textheight, keepaspectratio=true]{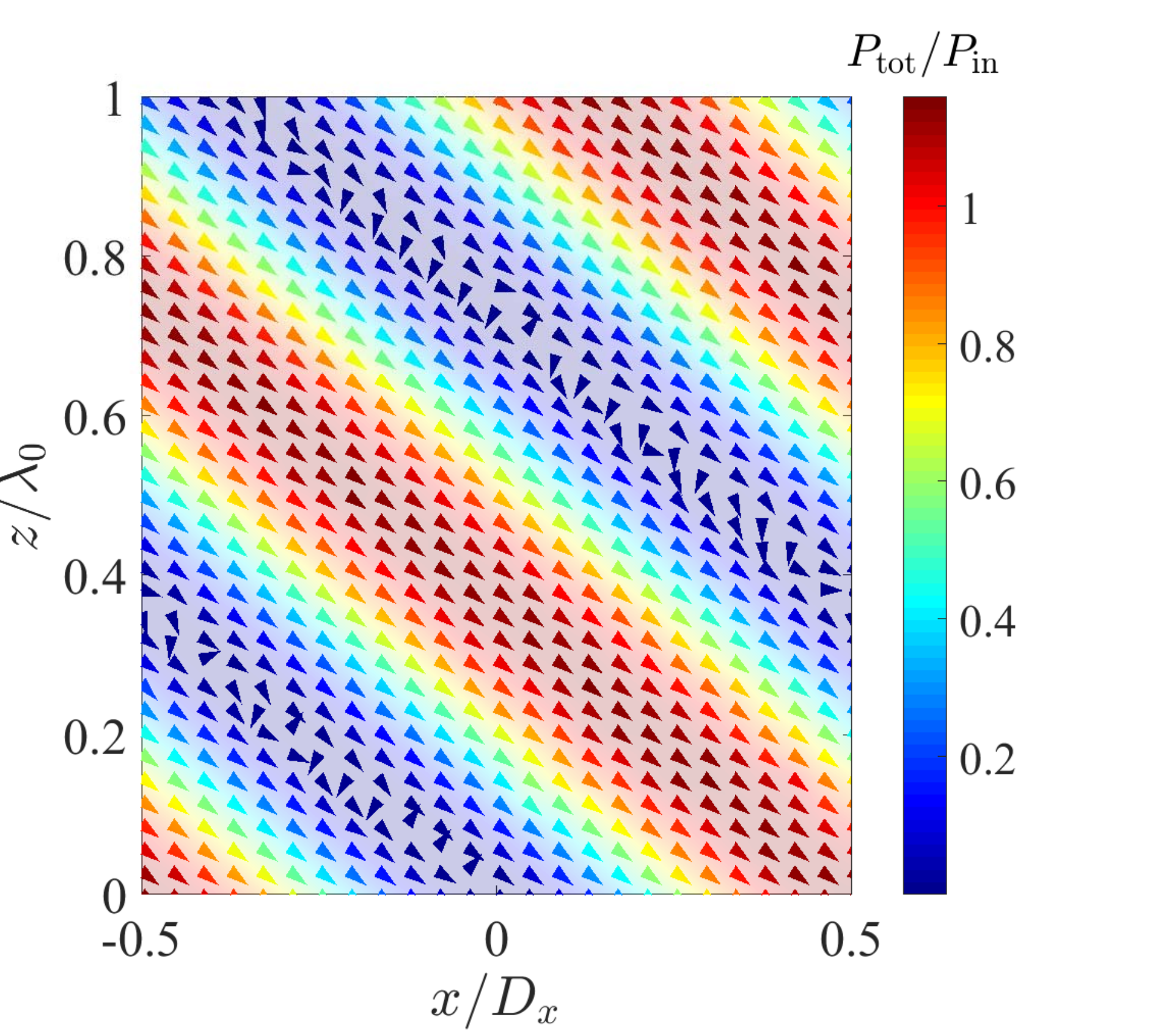}\label{fig:DiazRubioFIG2e}}
		\subfigure[]{\includegraphics[width = 0.22\textheight, keepaspectratio=true]{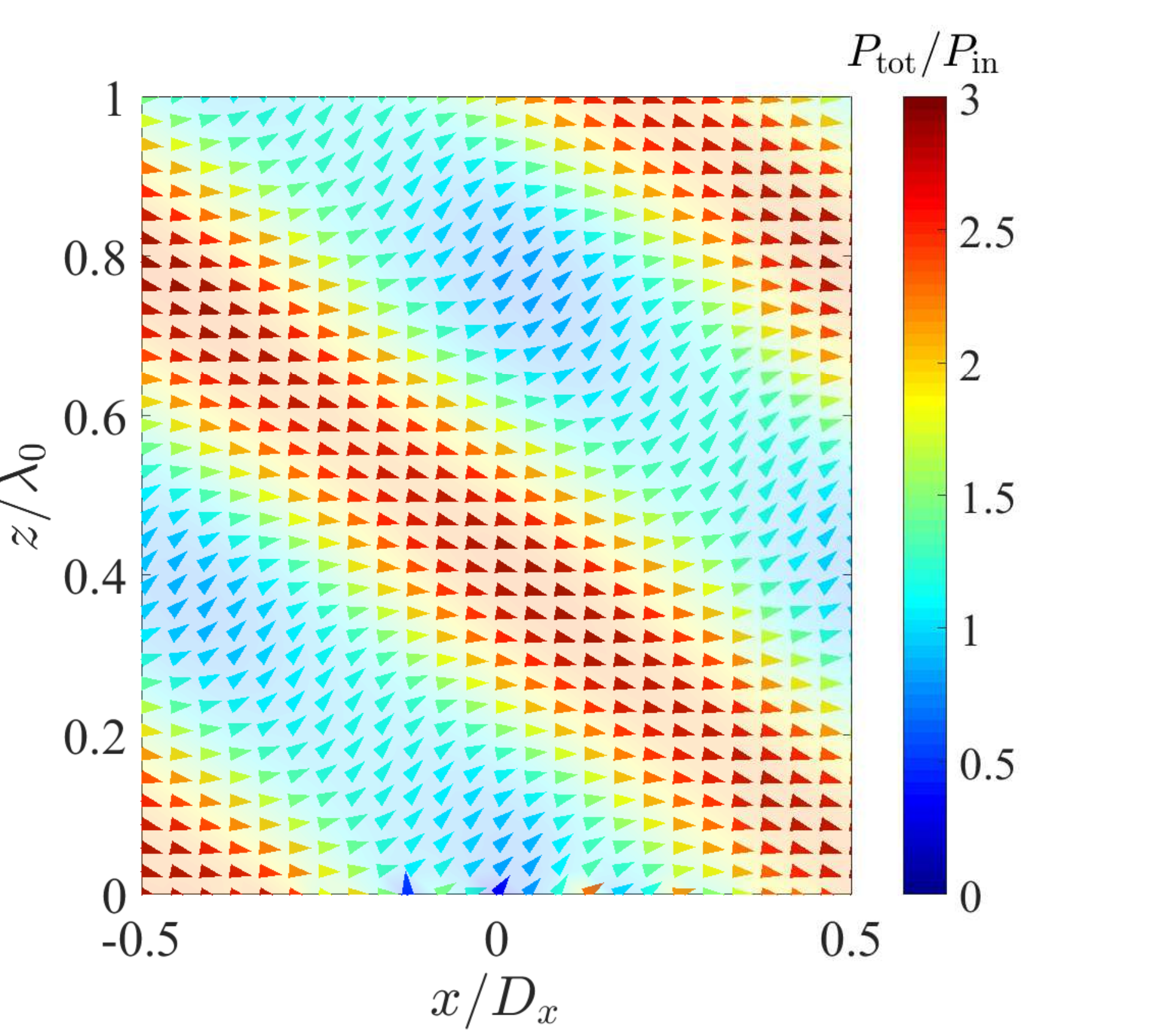}\label{fig:DiazRubioFIG2f}}
	\end{minipage} 
	\caption{Comparison between the different reflective metasurface proposals when $\theta_{\rm i}=0^\circ$ and $\theta_{\rm r}=70^\circ$. Conventional design: (a) Real part of the scattered electric field and (d) Total power density distribution. Lossy design: (b) Real part of the scattered electric field and (e) Total power density distribution. Active design: (c) Real part of the scattered electric field and (f) Total power density distribution.}\label{fig:DiazRubioFIG2}
\end{figure}

Let us look into this important feature in more detail. The period of the metasurface $D_{ x}$ is defined by requiring that the reflection phase is $2\pi$-periodic: $\Phi_{\rm r}(x)=\Phi_{\rm r}(x+D_{x})+2\pi$, and considering the phase shift in reflection dictated by the generalized Snell's law it can be expressed as  $D_x=\lambda/  \lvert{\sin\theta_{\rm r}-\sin\theta_{\rm i}}\rvert$, where $\lambda=2\pi/k_1$ is the wavelength. The period will define the directions where the reflected energy can flow. For example, considering normal incidence, $\theta_{\rm i}=0^\circ$, and $\theta_{\rm r}>30^\circ$, the energy only can be reflected as plane waves in three different directions (see Fig.~\ref{fig:DiazRubioFIG1_A}): the specular direction ($\theta_{\rm i}$), the desired direction ($\theta_{\rm r}$) and the symmetric direction ($-\theta_{\rm r}$). Numerical simulations have been done modeling a metasurface with the impedance boundary described by Eq.~(\ref{eq:conventional}). Particularly, the impedance boundary has been designed for reflecting the energy from $\theta_{\rm i}=0^\circ$ into $\theta_{\rm r}=70^\circ$. 
Figure~\ref{fig:DiazRubioFIG2a} shows the distribution of the real part of the scattered electric field, where we see an  interference pattern produced by these three plane waves. As previously stated, the energy is distributed into three directions producing a modulation of the real part of the Poynting vector, as it is shown in Fig.~\ref{fig:DiazRubioFIG2d}. 
It is important to notice that the amplitude of the reflected wave in the desired direction is higher than the incident amplitude $E_{\rm r}/E_{\rm i}=1.5$, so the metasurface is adding not only a linear phase shift as it was expected, but also changes the amplitude. This property agrees with the conclusion that more than two propagating plane waves must exist in the system. 
The efficiency of this conventional design is {$\xi_{\rm P}=P_{\rm r}/P_{\rm i}=0.76$} and the residuary energy is sent to the other directions. In this  definition of the power  efficiency, $P_{\rm r}$ and $P_{\rm i}$ are the amplitudes of the Poynting vector of the plane wave reflected in the desired direction and that of the incident plane wave, respectively.

Parasitic reflections can be suppressed by allowing power absorption in the metasurface. As is shown in \cite{Alu,synthesis}, a solution for the Maxwell equations where $E_{\rm r}=E_{\rm i}$ can be found and the corresponding surface impedance (for TE-polarized waves) reads 
\begin{equation} \label{eq:lossy}
Z_{\rm s}(x)= \eta_1\frac{1 + e^{j\Phi_{\rm r}(x)} }{\cos\theta_{\rm i} - \cos\theta_{\rm r}e^{j\Phi_{\rm r}(x)}}.
\end{equation}
The impedance given by Eq.~(\ref{eq:lossy}) is a complex number with some positive real part and the same period as in the conventional design (\ref{eq:conventional}). The real part in the surface impedance represents losses in the metasurface. Figure~\ref{fig:DiazRubioFIG2} shows the results of numerical simulations of a metasurface defined by Eq.~(\ref{eq:lossy}) when  $\theta_{\rm i}=0^\circ$ and $\theta_{\rm r}=70^\circ$. The real part of the scattered field is represented in Fig.~\ref{fig:DiazRubioFIG2b}. We can see that a perfect plane wave with the same amplitude than the incident wave is reflected into the desired direction.  
The real part of the Poynting vector is represented in Fig.~\ref{fig:DiazRubioFIG2e}, where we can see the power entering into the metasurface due to non-zero values of the real part of the impedance (the metasurface is lossy). The efficiency of this metasurface is $\xi_{\rm P}=P_{\rm r}/P_{\rm i}=0.34$ and the absorption is $A=1-P_{\rm r}/P_{\rm i}=0.66$.  
Figure~\ref{fig:DiazRubioFIG2e} shows that the magnitude of the power is modulated in the $x$-$z$ plane with a flat wavefront. From the comparison between the spatial distributions of the electric field and the Poynting vector, it is easy to see that the tilt angle defining this modulation, $\theta_{\rm power}$, is different from the direction of the reflected wave phase front, $\theta_{\rm r}$. The reader is referred to the Supplementary Materials for more information \cite{suppl}.

Very recently, it was shown that it is in principle possible to realize anomalously reflecting metasurfaces which operate perfectly, that is, without any parasitic reflections, scattering, and absorption \cite{synthesis,last}. In order to achieve perfect anomalous reflection,  the power carried in the desired direction must be equal to the power of the incident plane wave, so the amplitude of the reflected wave has to be $E_{\rm r}=E_{\rm i}\frac{\sqrt{\cos\theta_{\rm i}}}{\sqrt{\cos\theta_{\rm r}}}$ \cite{synthesis}. Considering this condition, the surface impedance can be written as
\begin{equation} \label{eq:active}
Z_{\rm s}(x)= \frac{\eta_1}{\displaystyle \sqrt{\cos\theta_{\rm i}\cos\theta_{\rm r}}}\frac{\sqrt{\cos\theta_{\rm r}} + \sqrt{\cos\theta_{\rm i}}\,e^{j\Phi_{\rm r}(x)} }{\displaystyle  \sqrt{\cos\theta_{\rm i}} -\sqrt{\cos\theta_{\rm r}} \, e^{j\Phi_{\rm r}(x)}}.
\end{equation}
The analysis of this expression shows that the input impedance is a complex number, whose period is equal to the conventional design (\ref{eq:conventional}) and the lossy design (\ref{eq:lossy}). 
The real part of the input impedance periodically takes positive (corresponding to loss) and negative (gain) values. The power which passes through the input surface in the ``lossy'' regions is  re-radiated from the ``active'' regions so that the overall metasurface response is lossless. 
Figures~\ref{fig:DiazRubioFIG2c} and \ref{fig:DiazRubioFIG2f} show numerical simulations for this design when  $\theta_{\rm i}=0^\circ$ and $\theta_{\rm r}=70^\circ$. The real part of the scattered electric field is represented in Fig.~\ref{fig:DiazRubioFIG2c}. In this case, the ratio between the scattered and incident fields $E_{\rm r}/E_{\rm i}=1.7$ fulfils the condition for the power conservation previously mentioned, and the power efficiency equals 100\%. 
Figure~\ref{fig:DiazRubioFIG2f} shows the power flow where one can visualize this behaviour.  
Actual implementations of such perfect anomalous reflectors can be done by using the following approaches: (i) Including active and lossy elements in the metasurfaces \cite{Alu} as it is dictated by Eq.~(\ref{eq:active}). However, the use of active elements is usually not desirable for practical reasons and leads to potentially unstable structures; (ii) Using auxiliary evanescent fields for avoiding the modulation of the normal component of the Poynting vector at the metasurface, as it was theoretically demonstrated in \cite{last} for penetrable metasurfaces (metasurfaces allowing fields at both sides). In these theoretical designs, by engineering the evanescent fields excited at the metasurface, one can ensure the local power conversation at each point of the metasurface; (iii) Creating passive but non-local metasurfaces and engineer the interactions between the constitutive elements, i.e. the effects of evanescent fields  generated in the array, for allowing proper energy channeling along the metasurface. In this last scenario, the cell-averaged normal component of the Poynting vector equals to zero, while  the local behavior appears either lossy or active.

Inspired by the physical principle of leaky-wave antennas, we introduce a new approach based on the modulation of the reactive impedance of the metasurface.
We demonstrate that a perfect anomalous reflector can be realized as a simple metal pattern on a thin grounded dielectric slab. In this scenario, engineered modulations of the surface reactance ensure the required non-local reflections, which are eventually perfectly launched only in the desired direction. We present and clarify the main conceptual differences between the local and non-local approaches in the use of evanescent fields for realizing perfect anomalous reflectors.  Based on the developed theory we design, manufacture, and experimentally study an example metareflector which reflects a normally incident plane wave into a plane wave at the tilt angle $\theta_{\rm r}=70^\circ$. The experimental results  confirm that this first prototype of metasurfaces for perfect control of reflections operates according to expectations: the realized power efficiency is close to 100\%, limited only by dissipation in the metal patches and the dielectric substrate. Parasitic reflections into the specular direction and other directions are seen to be negligible. The proposed simple topology  allows  cheap mass-production manufacturing based on the conventional printed circuit board technology (microwave frequencies) or various lithography techniques (terahertz and beyond). 

\section{Results}
\subsection{Theoretical local design of a lossless perfect anomalous reflector}

\begin{figure}
	\centering
	\begin{minipage}{\linewidth}
		\subfigure[]{\includegraphics[width = 0.22\textheight, keepaspectratio=true]{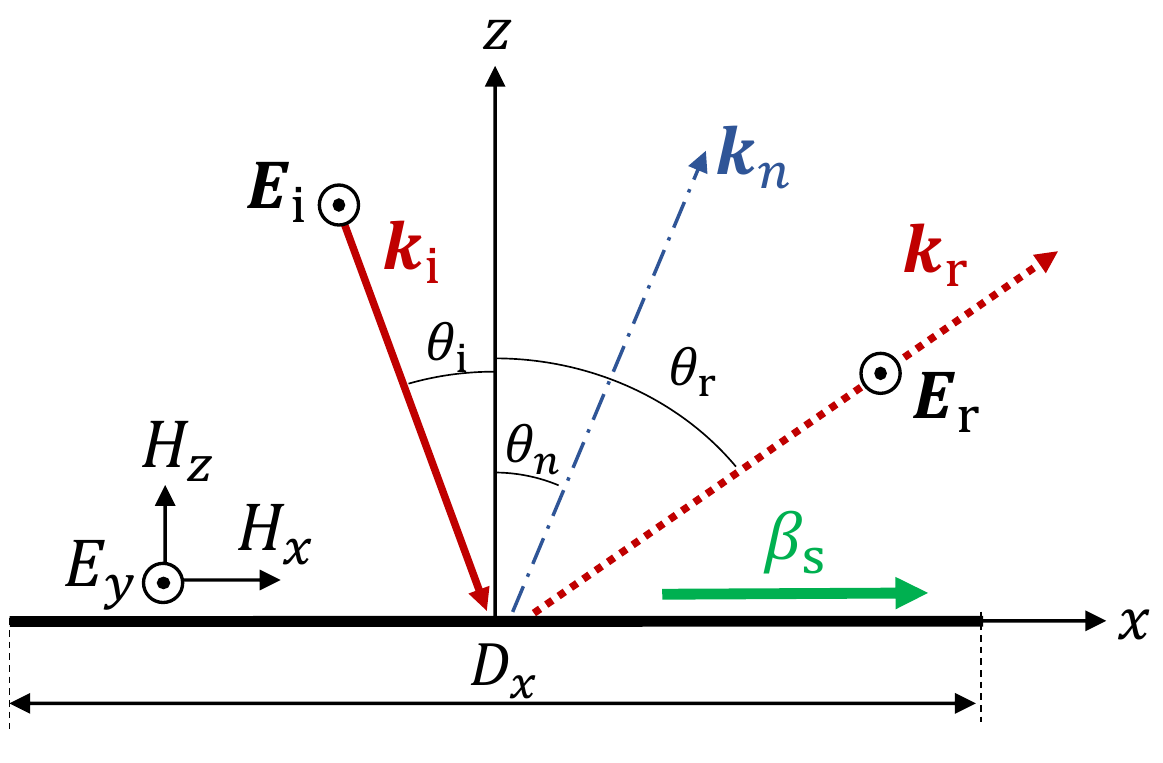}\label{fig:DiazRubioFIG3a}}
		\subfigure[]{\includegraphics[width = 0.22\textheight, keepaspectratio=true]{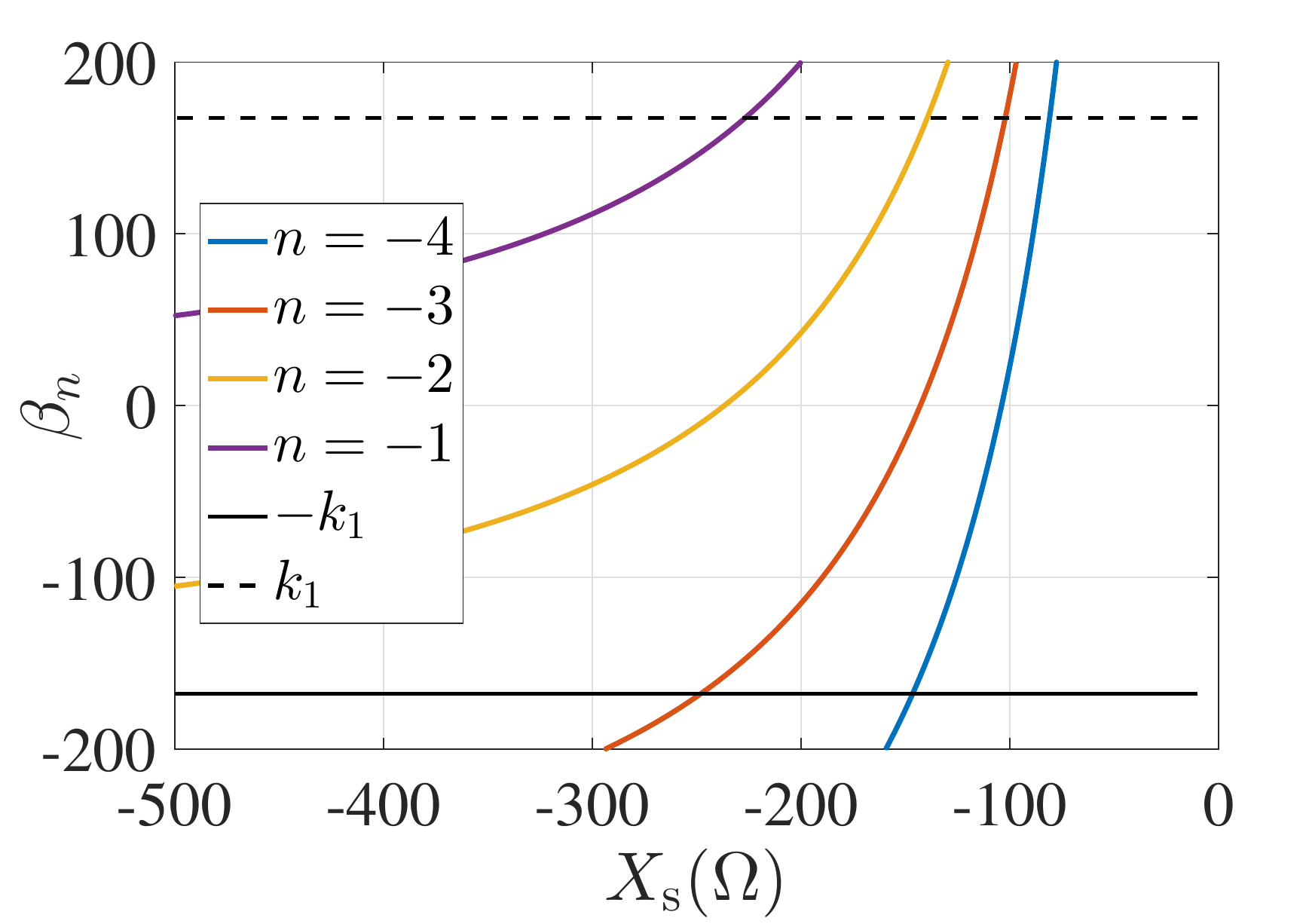}\label{fig:DiazRubioFIG3b}}
		\subfigure[]{\includegraphics[width = 0.22\textheight, keepaspectratio=true]{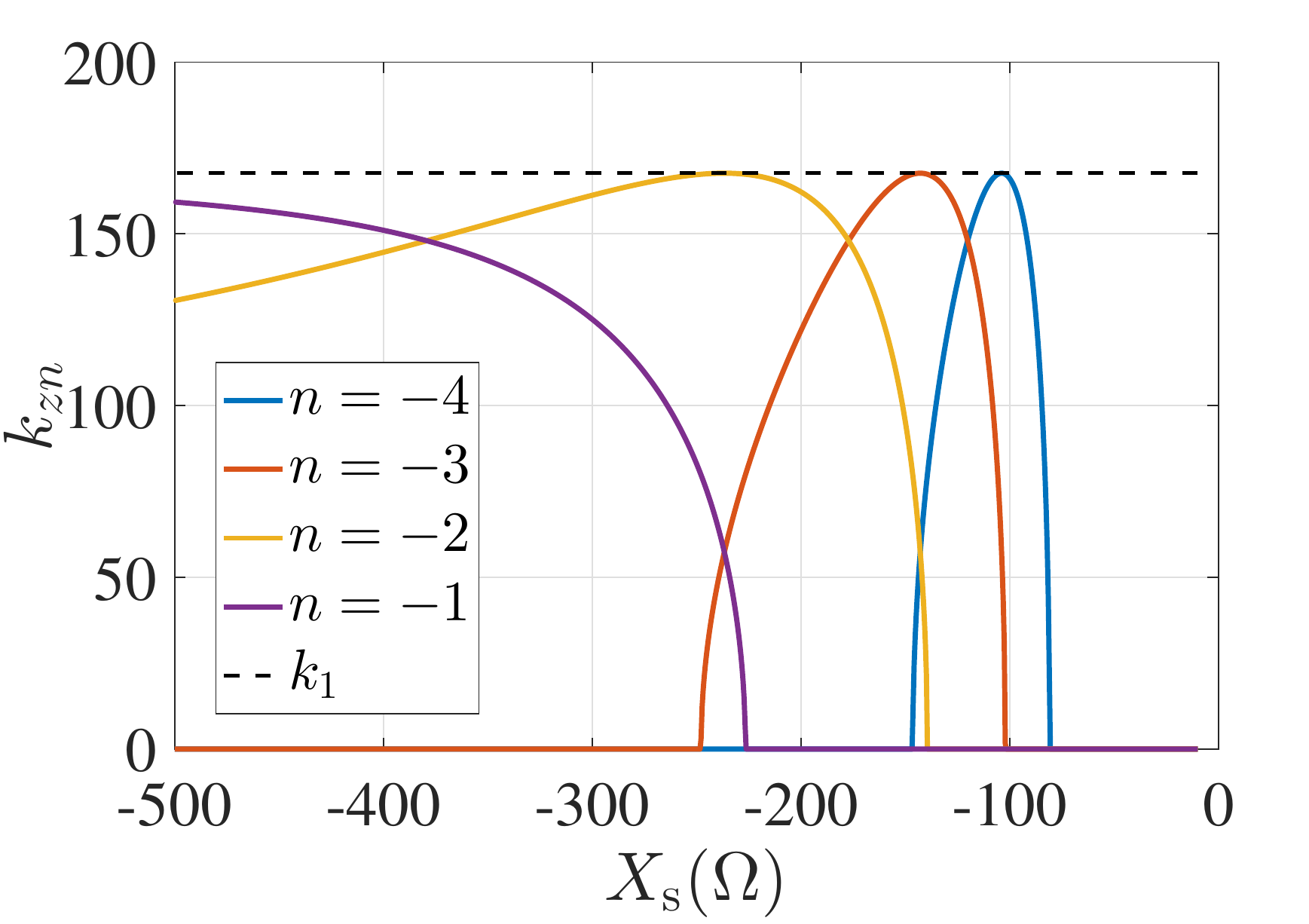}\label{fig:DiazRubioFIG3c}}
	\end{minipage}
	\begin{minipage}{\linewidth}
		\subfigure[]{\includegraphics[width = 0.22\textheight, keepaspectratio=true]{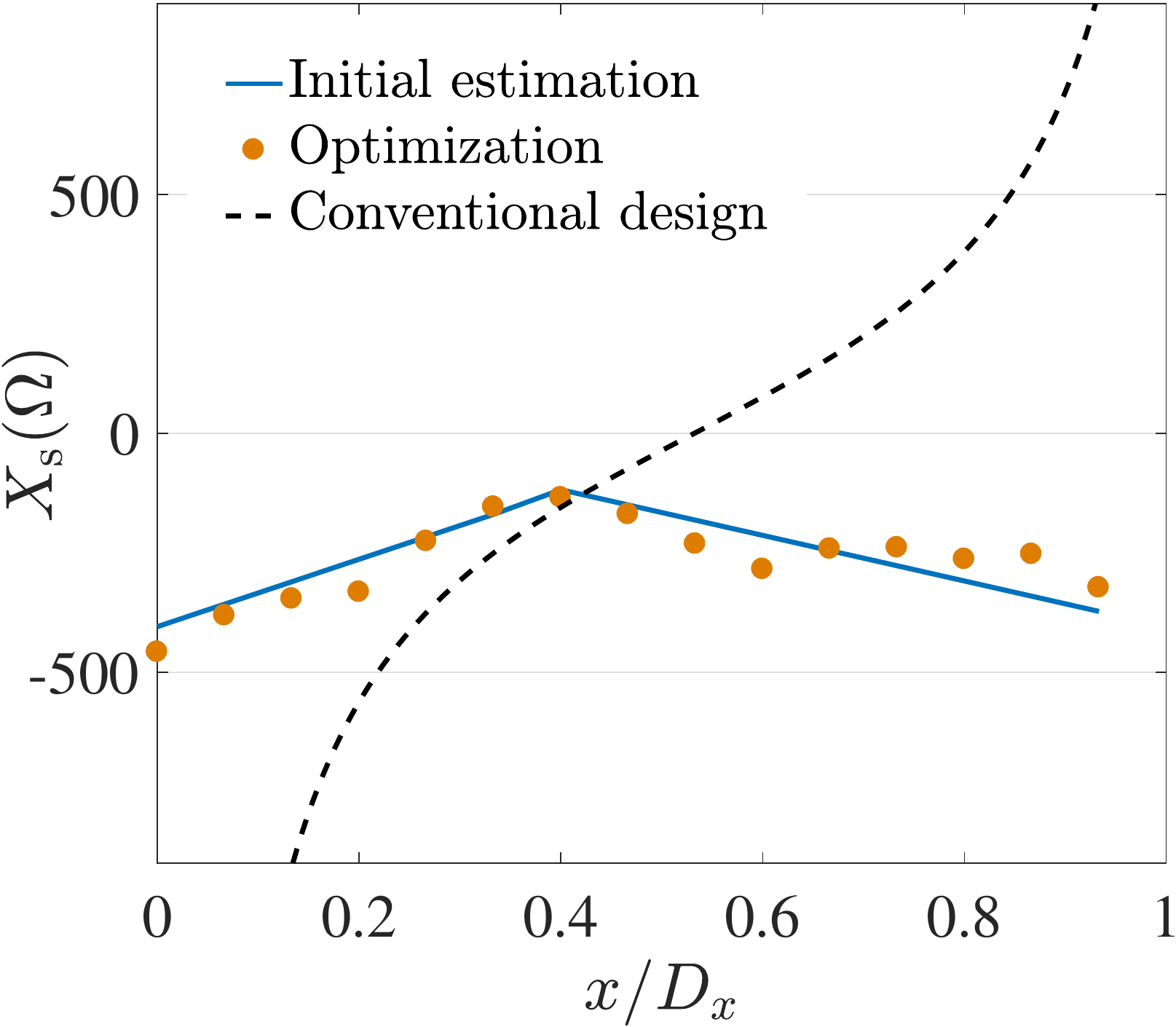}\label{fig:DiazRubioFIG3d}}
		\subfigure[]{\includegraphics[width = 0.22\textheight, keepaspectratio=true]{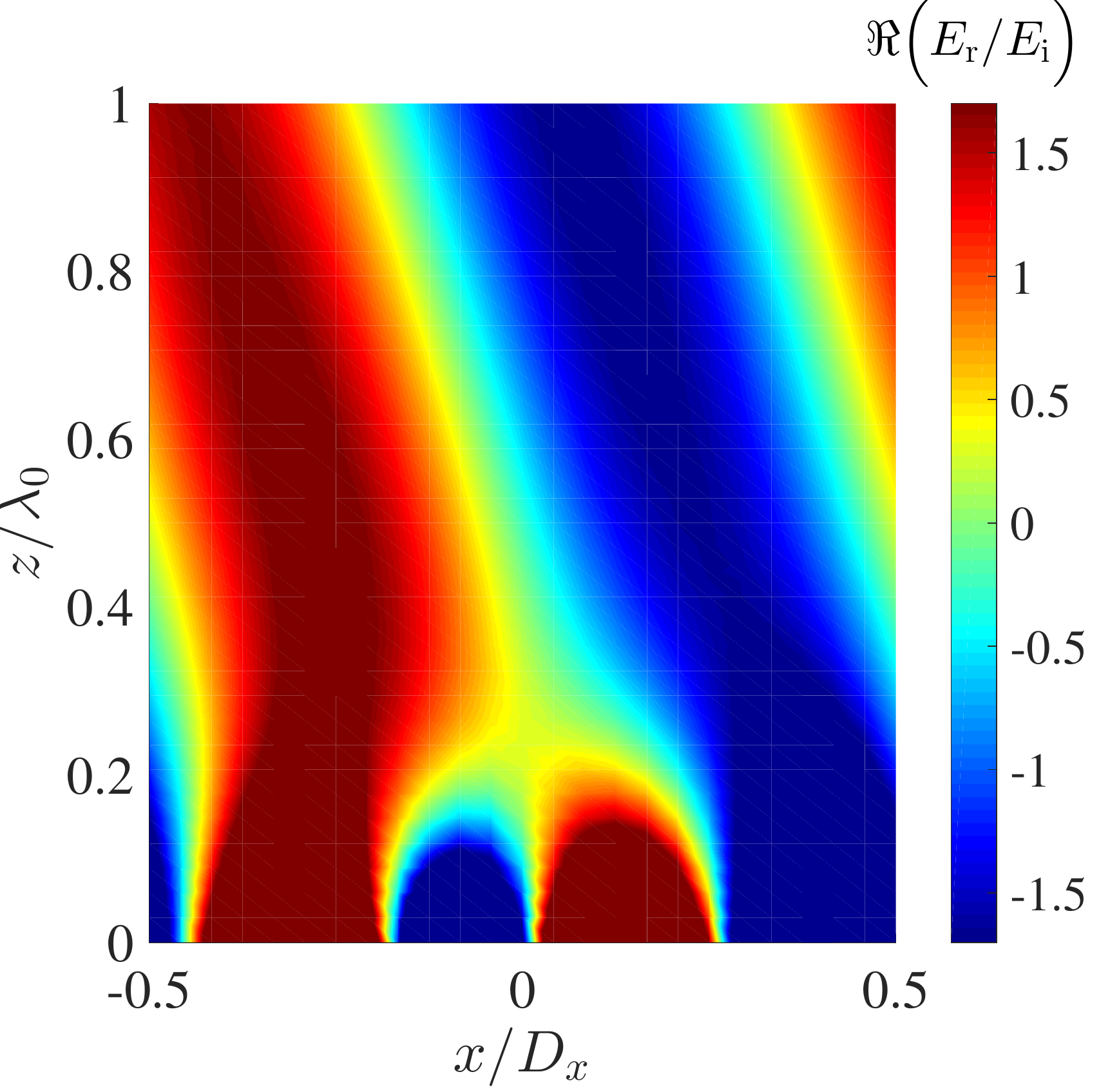}\label{fig:DiazRubioFIG3e}}
		\subfigure[]{\includegraphics[width = 0.22\textheight, keepaspectratio=true]{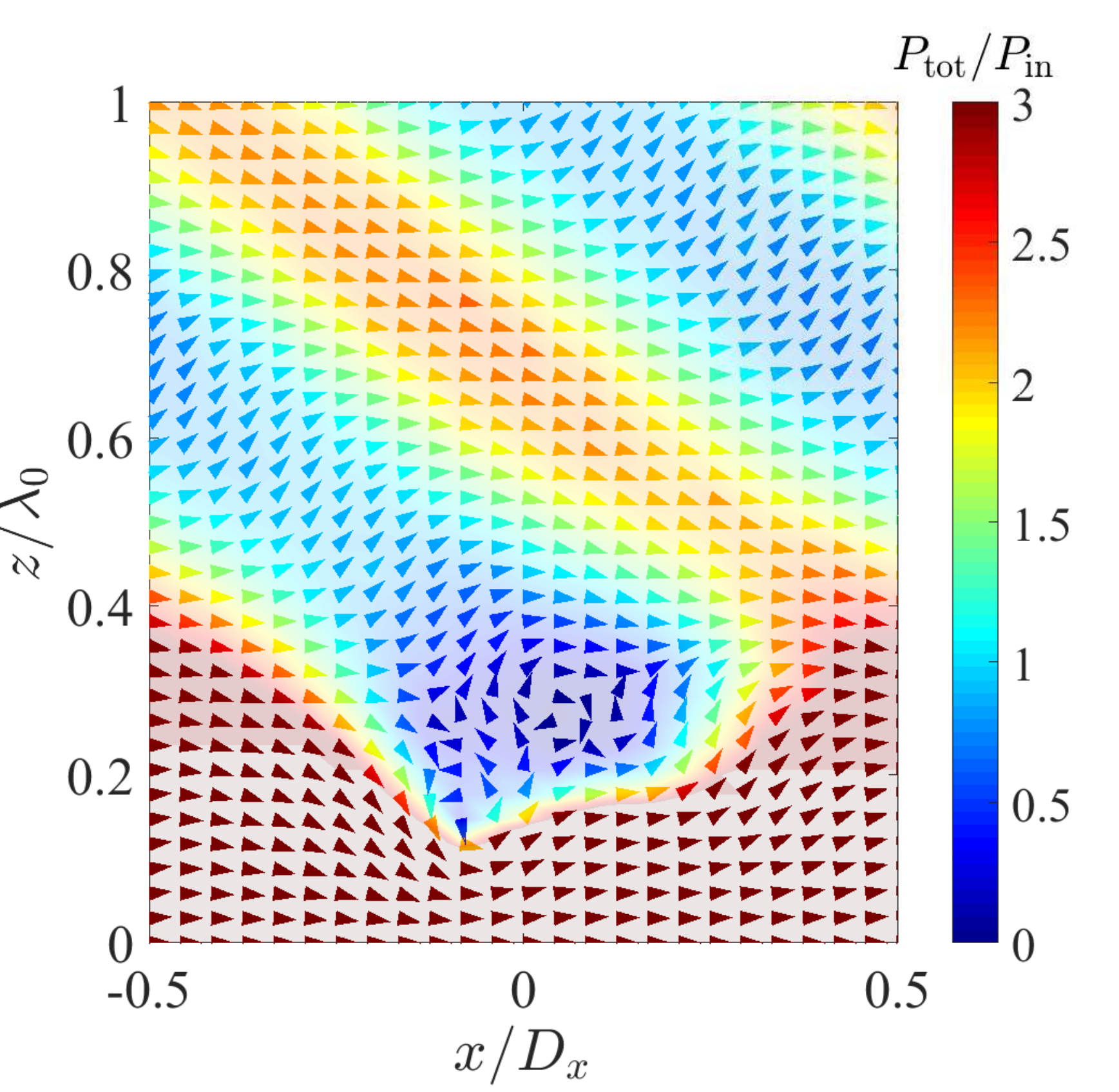}\label{fig:DiazRubioFIG3f}}
	\end{minipage} 
	\caption{(a) Schematic representation of the proposed design methodology for the perfect reflecting metasurface based on the leaky wave antenna behaviour with a modulation in the impedance profile. (b) Tangential and  (c) vertical  wavenumbers introduced in system due to  a periodic perturbation with period  $D_x=\lambda/\lvert \sin \theta_{\rm i}-\sin \theta_{\rm r} \rvert$ as a function of the surface impedance $Z_{\rm s}=jX_{\rm s}$. (d) Surface impedance of the metasurface when $\theta_{\rm r}=70^\circ$ and $\theta_{\rm i}=0^\circ$. Blue line represents the initial estimation of the surface reactance  (Eq.~(\ref{eq:impedance})) and red dots represent the optimized values of the surface reactance  on the discretized surface. Dashed line represents the impedance of conventional design described by Eq. \ref{eq:conventional}. Numerical results of a perfect reflectarray based on the linearly-modulated leaky-wave antenna structure: (e) Real part of the scattered electric field. (f) Total power density flow distribution.}\label{fig:DiazRubioFIG3}
\end{figure}

First, we consider a surface-impedance model of the metasurface and study the problem by using locally responding impedance boundaries.
We think that the most reasonable approach to obtain the  impedance boundary which realizes  perfectly reflecting lossless reflectors is to use inhomogeneous leaky-wave surfaces, generalizing the concept of leaky-wave antennas.
Indeed, leaky-wave antennas are formed by perturbed waveguides or transmission lines, naturally offering the needed channel for transporting power from receiving areas to transmitting areas. Even more importantly, tuning the inhomogeneity profiles along the antenna surface it appears possible to realize regimes where the receiving areas receive power predominantly from the illumination direction while the radiating areas send the waves  predominantly in the desired direction of anomalous reflection. Thus, tuning the phase synchronism of free-space waves and leaky waves on the surface we can realize directive power transfer along the surface without using any non-reciprocal elements.

We start the study considering the surface-impedance model of a leaky-wave structure. Within this model, lossless reflectors are described by a  reactive surface impedance $Z_{\rm s0}$. In order to allow propagation of a surface wave of the considered TE polarization along the surface, $E_y=E_0e^{-j\beta_{\rm s}x-\alpha z}$, we demand that the reactance is negative (capacitive) at every point of the surface $Z_{\rm s0}=E_y/H_x=-j\omega \mu_0/\alpha$. The tangential wavenumber of the surface wave can be found using the relation $\beta_{\rm s}^2=k_1^2+\alpha^2$. Thus, the tangential wavenumber can be written as
\e \beta_{\rm s}=k_1\sqrt{1-{\eta_1^2\over Z_{\rm s0}^2}}. \f
In order to couple the surface wave to free space,  the reactance should be non-uniform over the reflector surface. In conventional leaky-wave antennas, where the goal is to launch a wave in a specific direction, the surface is periodically perturbed. This perturbation generates spatial harmonics whose tangential wave number can be expressed as
\e \beta_{n}=\beta_{\rm s}+n\frac{2\pi}{D_x}, \f
where $n=\pm1,\pm2...$. The corresponding vertical wavenumbers, $k_{zn}$, can be found using 
\e k_{zn}=\sqrt{ {{k_1}^2-{\beta_{n}}^2} }.\f
Knowing  the vertical and tangential wavenumbers, the direction of the propagating waves is calculated using
\e \sin\theta_{n}=\frac{\beta_{n}}{k_1}=\sqrt{1-{\eta_1^2\over Z_{\rm s0}^2}}+n\sin\theta_{\rm r}.\f
Figures~\ref{fig:DiazRubioFIG3b} and \ref{fig:DiazRubioFIG3c} represent the tangential and vertical wavenumbers for different values of the surface impedance $Z_{\rm s0}=jX_{\rm s0}$.

The impedance surface should be chosen in order to couple the energy to free space through leaky waves. For example, if we choose $X_{\rm s0}=-124\ \Omega$,  only the harmonics $n=-3$ and $n=-4$ propagate in free space. Particularly, for  $X_{\rm s0}=-124\ \Omega$ the waves propagate in the directions specified by $\theta_{\rm -3}=22.3^\circ$ and $\theta_{\rm -4}=-34^\circ$.  Once we have determined the surface impedance that allows surface-wave propagation and coupling with free-space waves, the reactance should be modulated. Conventional periodical modulations (usually with a sinusoidal profile) provide coupling to the waves along these two directions (in this example). But as we saw above, to realize perfect anomalous reflection, in the areas where the surface should receive power from the incident field, the wave along the surface should be in phase synchronism with the incident plane wave. In contrast, in the areas where the energy should be launched into the desired direction,  the synchronization  should hold for the reflected plane wave. 

To realize such operation, we propose to modulate the reflection phase linearly, using the generalized reflection law separately for these two parts of the metasurface period. 
The required derivative of the local phase can be estimated by considering  the additional  effective ``momentum'' along the surface, in analogy with the generalized law of reflection. The required linear phase dependence can be written as
\e \Phi_{\rm r}(x)=
	\begin{cases} 
		  \Delta\theta_{\rm r}\,k_1 x- \Phi_{\rm 0} & 0 \leq x< x_1 \\
		  \Delta\theta_{\rm i}\, k_1 (x-D_x) -\Phi_{\rm 0} & x_1\leq x < D_x ,
		     \end{cases}
\f
where $\Delta\theta_{\rm r}=\sin\theta_{\rm r}-\sin\theta_{n}$ and $\Delta\theta_{\rm i}=\sin\theta_{\rm i}-\sin\theta_{n}$. In our particular example, the derivative has the opposite sign in the two parts of the surface period (shifting the angle $\theta_{\rm -3}=22.3^\circ$ to zero and $70^\circ$, respectively), and  $x_1=D_x\frac{\sin\theta_{\rm n}-\sin\theta_{\rm i}}{\sin\theta_{\rm r}-\sin\theta_{\rm i}}$ is the point where the derivative of the reflection phase changes from positive to negative. The initial phase shift reads  
\begin{equation}
\Phi_{\rm 0}=2\pi\frac{(\sin\theta_{\rm r}-\sin\theta_{n})(\sin\theta_{n}-\sin\theta_{\rm i})}{(\sin\theta_{\rm r}-\sin\theta_{\rm i})^2}-2\arctan{\frac{X_{\rm s0}}{\eta_1}}.
\end{equation}
Notice that ${\Phi_{\rm r}(x_1)}=2\arctan{\frac{X_{\rm s0}}{\eta_1}}$ corresponds to the phase of the reflection coefficient for the homogeneous surface impedance  $Z_{\rm s0}=jX_{\rm s0}$. The relation between the phase gradient and the modulated surface impedance can be found through the reflection coefficient
\begin{equation} \label{eq:phase}
\Phi_{\rm r}(x)=2\arctan{\frac{X_{\rm s}(x)}{\eta_1}}\approx 2\frac{X_{\rm s}(x)}{\eta_1},
\end{equation}
where the approximation holds for small values of $X_{\rm s}/\eta_1$. 
Using Eq.~(\ref{eq:phase}), we can estimate the modulations needed for the metasurface input impedance using the following expression: 
\begin{equation} \label{eq:impedance}
 X_{\rm s}(x)=	
	\begin{cases} 
		\frac{\eta_1}{2}(k_1\Delta\theta_{\rm r}\, x- \Phi_{\rm 0})& 0 \leq x< x_1 \\
		\frac{\eta_1}{2}(k_1\Delta\theta_{\rm i}\,  (x-D_x) -\Phi_{\rm 0}) & x_1\leq x < D_x.
    \end{cases}
\end{equation}

The blue line in Fig.~\ref{fig:DiazRubioFIG3d} represents the surface reactance profile defined by Eq.~(\ref{eq:impedance}) when $X_{\rm s0}=-124\ \Omega$, $n=-3$, $\theta_{\rm i}=0^\circ$, and $\theta_{\rm r}=70^\circ$. Obviously, this analytical estimation of the perfect reactance profile is rather approximate, because we make use of the homogeneous reactance model in case when the assumption that the metasurface is uniform on the wavelength scale is not properly justified. For this reason next we do numerical optimization of the surface reactance, setting these estimations of the required reactance profile as the initial guess and using 15 elements for the discretization as a piece-wise homogeneous reactive impedance boundary. As a result, we find the profile shown in Fig.~\ref{fig:DiazRubioFIG3d} with the red dots (one dot corresponds to one homogeneous reactive-surface element). As expected from the above theory, it is everywhere capacitive, growing in one half of the period and decaying inside the other half. The differences with the analytically predicted values are due to the approximations in the models and are caused mainly because of the periodic conditions imposed over the unit cell and the piece-wise constant numerical model of the surface that force generation of more complex evanescent field structure than assumed in the analytical analysis.

Numerical simulations of the corresponding field and the Poynting vector distributions are shown in Fig.~\ref{fig:DiazRubioFIG3e} and \ref{fig:DiazRubioFIG3f}. We clearly see that a surface wave propagating along the surface is indeed formed. If we now define a reference plane above the volume filled by the surface-mode fields (which is our ``input port'' to the metasurface structure) and look at the input impedance there, we see that it indeed satisfies  the requirements of perfect operation: it is a complex value given by Eq.~(\ref{eq:active}), and the real part is properly varying, emulating ``loss'' where the power is received by the leaky-wave structure and ``gain'' in the areas where it is launched back.  

\begin {table*}
\begin{tabular}{|L{0.3\linewidth}| C{0.11\linewidth}| C{0.11\linewidth}| C{0.11\linewidth}| C{0.12\linewidth}|  C{0.1\linewidth}| }
	\hline
	& $\theta_{\rm i}$             & $\theta_{\rm r}$            & $-\theta_{\rm r}$  & Absorption   & Efficiency \\ \hline
	Generalized reflection law. Equation (\ref{eq:conventional}) &     0.24/ 0.06   &       1.50/ 0.76      &   0.73/ 0.18  &       0.00      &  75.7 $\%$ \\ \hline
	Lossy design dictated by Equation (\ref{eq:lossy})    &       0.00/0.00      &  1.00/0.34     &           0.00/0.00     &       0.66         &  34.0 $\%$ \\                  \hline
	"Active-lossy" design dictated by Equation (\ref{eq:active})   &         0.03/0.00            &     1.77/ 1.04            &         0.11/0.00  &       0.00     &  104.4$\%$ \\                  \hline
	Inhomogeneous leaky-wave antennas introduced in this paper.      &          0.04/0.00                &           1.70/ 0.99                &    0.03/0.00                      &       0.00      &  99.7$\%$ \\                  \hline
	Implementation with metal patches (lossy materials)    &            0.04/0.00              &        1.66/0.94                &            0.03/0.00              &       0.06      &  94.0$\%$   \\                \hline
\end{tabular}
\caption {Numerical results and comparison between the different design  possibilities for reflectarrrays. Amplitude/power of waves sent into the respective directions, absorption coefficient and power efficiency.}\label{tab:Comparison}
\end {table*}

Table~\ref{tab:Comparison} summarizes the field amplitudes and the power sent into the three directions ($\theta_{\rm i}$, $\theta_{\rm r}$, $-\theta_{\rm r}$) when the metasurface is illuminated by a normally incident plane wave for the different design options. It is clear that the inhomogeneous leaky-wave antenna design promises perfect performance.
\begin{figure*}
	\centering
	\begin{minipage}{0.41\linewidth}
		\subfigure[]{\includegraphics[width = 1\linewidth, keepaspectratio=true]{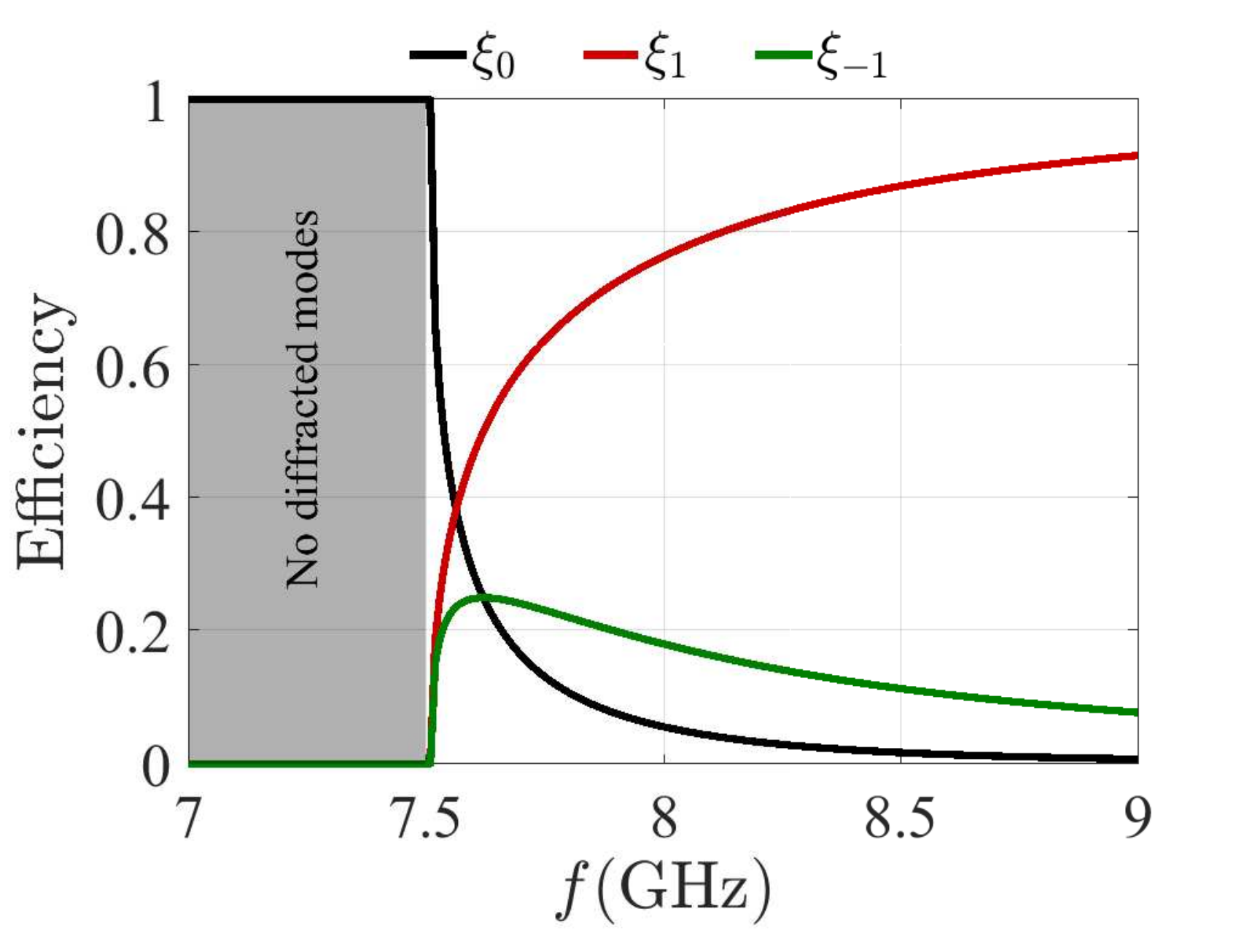}\label{fig:DiazRubioFIGr4a}}
		\subfigure[]{\includegraphics[width = 1\linewidth, keepaspectratio=true]{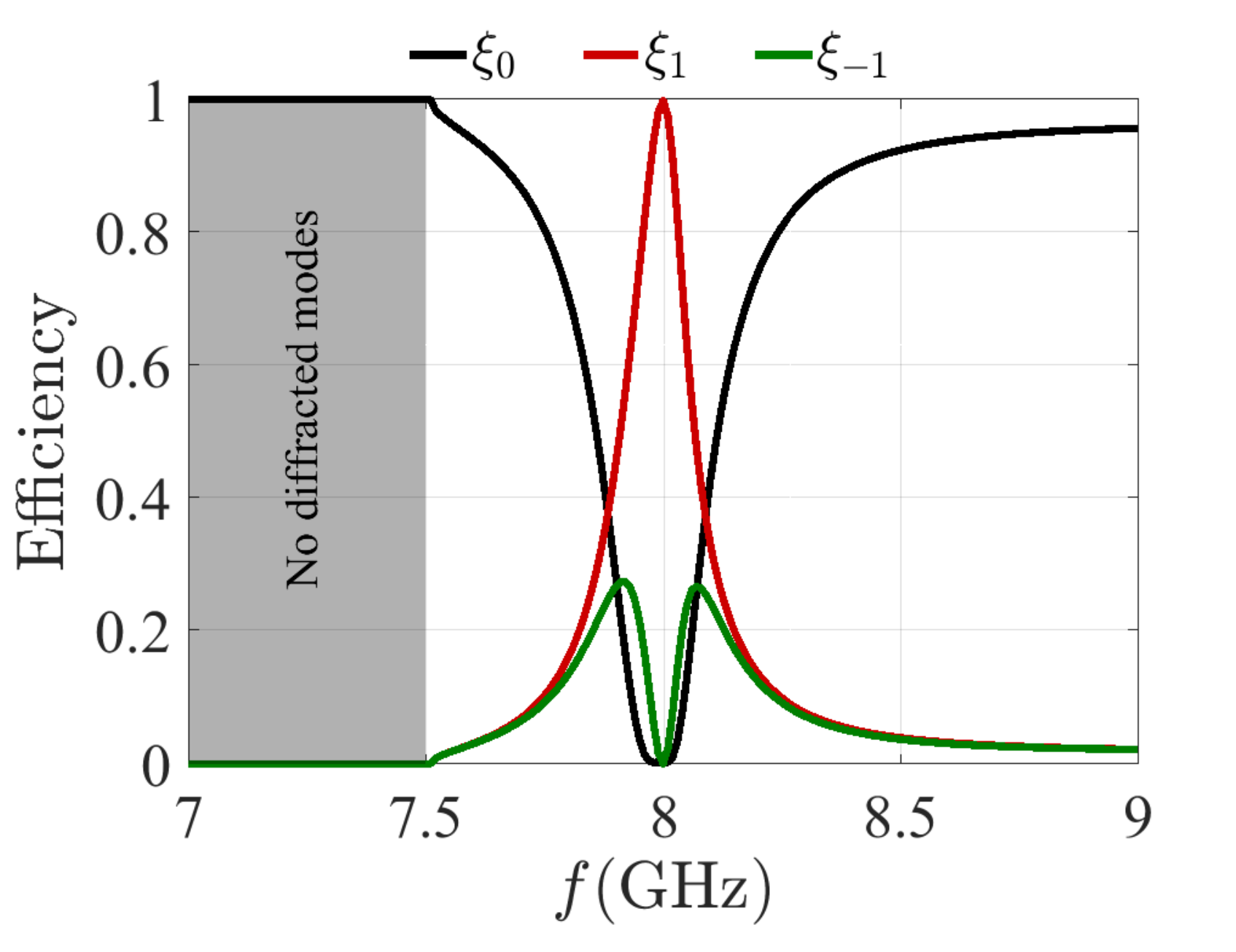}\label{fig:DiazRubioFIGr4b}}
	\end{minipage}  
	\begin{minipage}{0.5\linewidth}
		\subfigure[]{\includegraphics[width = 1\linewidth, keepaspectratio=true]{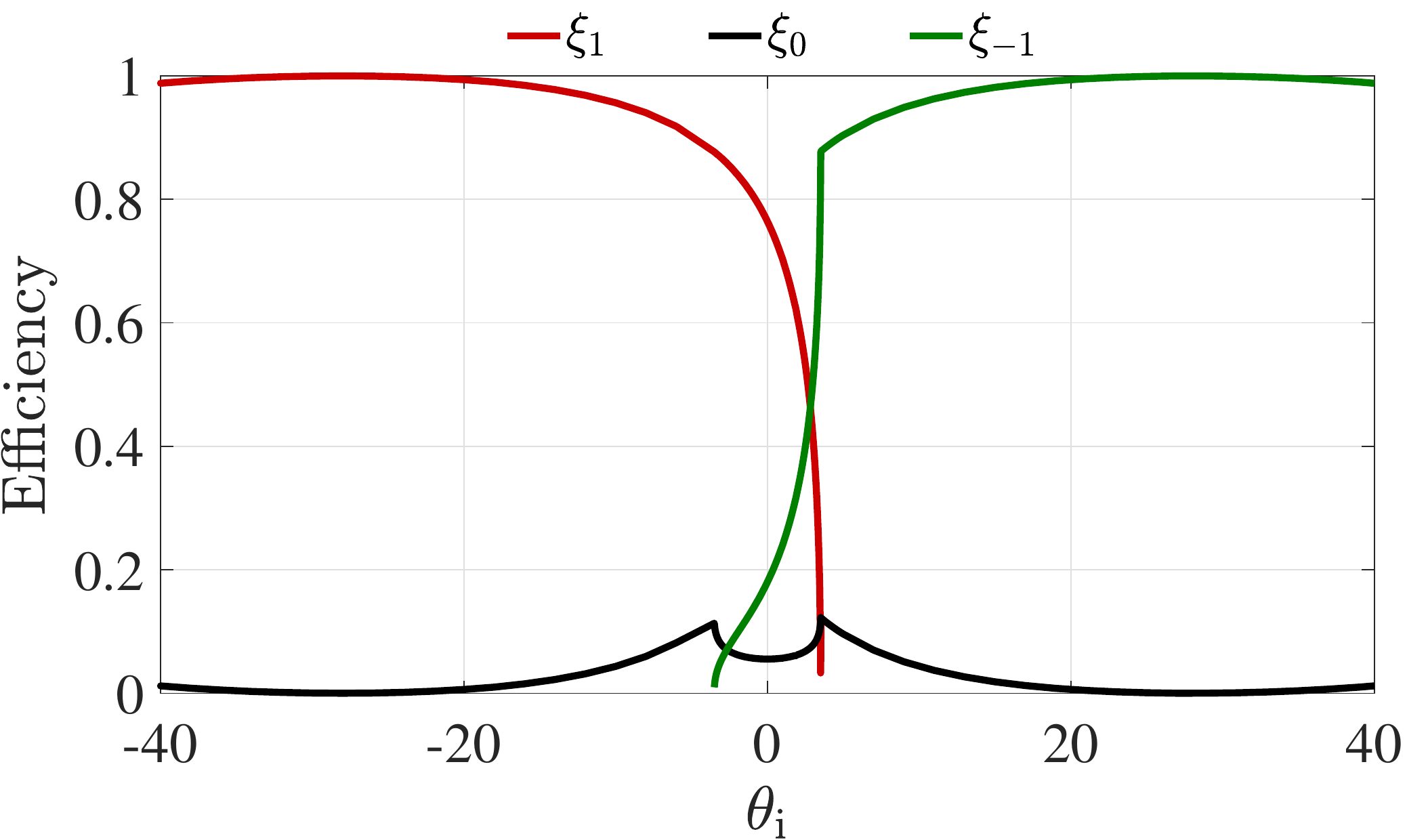}\label{fig:DiazRubioFIGr4c}}
		\subfigure[]{\includegraphics[width = 1\linewidth, keepaspectratio=true]{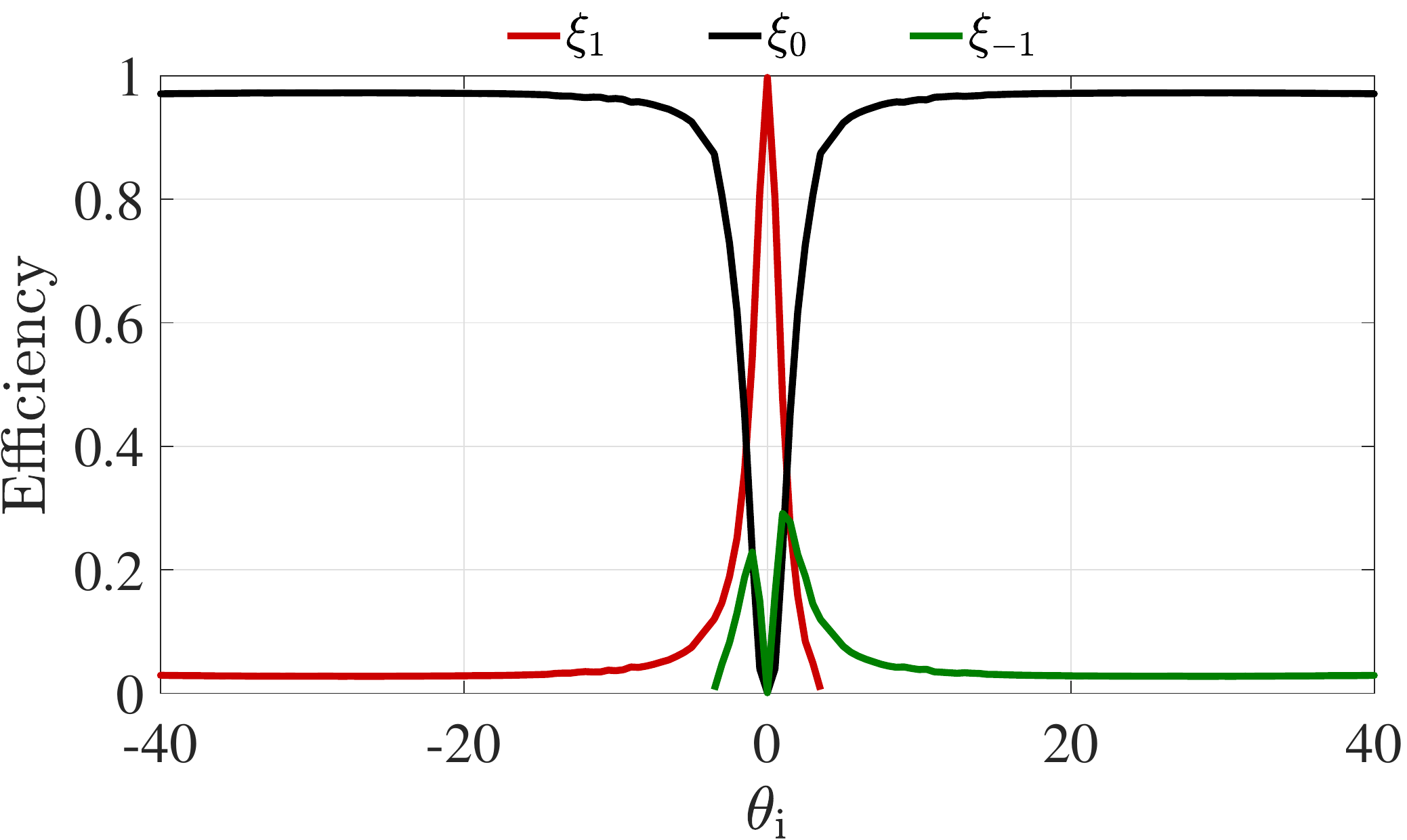}\label{fig:DiazRubioFIGr4d}}
	\end{minipage}
	\caption{Bandwidth comparison between anomalous reflectors based on  (a), (c) conventional design [Eq. (\ref{eq:conventional})] and  on  (b), (d)  inhomogeneous leaky-wave antenna [see Fig. \ref{fig:DiazRubioFIG3d}]. Both metasurfaces are designed for producing anomalous reflection from $0^\circ$ to $70^\circ$ at $8~{\rm GHz}$ and modeled as an inhomogeneous impedance boundary.  (a) and (b) Comparison between wavelength bandwidths. (c) and (d) Comparison between angular bandwidths.} \label{fig:DiazRubioFIG}
\end{figure*}

To complete the study of this metasurface, we consider the frequency bandwidth, Fig.~\ref{fig:DiazRubioFIGr4a},  and give a comparison with the conventional designs, Fig.~\ref{fig:DiazRubioFIGr4b}.  In both cases, below $7.5~\rm{GHz}$ no diffracted modes are allowed in the system and consequently all the energy is reflected back in the normal direction. In the range of frequencies between $7.5$ and $9~\rm{GHz}$, the behavior of the two designs  is different. At these frequencies three diffraction modes are allowed ($n=0,\pm1$), and the efficiency of each mode $\xi_n$ is calculated as the ratio between the power reflected into the $n$-th mode and the incident power. The values at $8~\rm{GHz}$ correspond to the values reported in Table~\ref{tab:Comparison}.
The conventional design based on a $2\pi$ linear phase gradient along the period shows a broadband response in the assumption that Eq.~(\ref{eq:conventional}) is exactly satisfied at all frequencies [see Fig.~\ref{fig:DiazRubioFIGr4a}]. The efficiency of the $n=1$ mode increases due to a reduction of the reflection angle. Considering the relation between the frequency and the reflection angle, this behavior is similar to a dispersive prism where waves of different   frequencies are sent into different directions (see Video I  in supplementary material).  On the other hand, the design based on the inhomogeneous leaky-wave antenna shows a completely different behavior [see Fig.~\ref{fig:DiazRubioFIGr4b}]. In this case, the anomalous reflection is a relatively narrow-band phenomena ($\xi_1>0.5$ from $7.9$ to $8.09~\rm{GHz}$), and  the metasurface acts as a mirror for other frequencies. This feature can be useful for narrow filtering or monochromatic emitters. It is important to notice that in both scenarios the model assumes that the boundaries are not dispersive with respect to the frequency or the incidence angle. Thus, the frequency dispersion of the response is caused only by the properties of the phase gradient. In  any physical implementation the response will be modified due to the frequency and spatial dispersion of the metasurface structure.

Figures~\ref{fig:DiazRubioFIGr4c} and \ref{fig:DiazRubioFIGr4d} compare the angular bandwidth of both designs, i.e, the power sent in each diffracted mode as a function of  the incidence angle. In this  study the incidence angle varies from $-40 ^\circ$ to $40 ^\circ$. In this range, there are three different regions: from  $-3.5 ^\circ$ to $3.5 ^\circ$, where three diffraction modes are allowed ($n=0,\pm1$); from $-40 ^\circ$ to $-3.5 ^\circ$ with only two diffracted modes ($n=0,1$); from $3.5 ^\circ$ to $40 ^\circ$, also with two diffraction modes ($n=0,-1$). Similarly to the results for the frequency response, the behaviors of the two designs are completely different. The design introduced in this work has a sharp response with respect to the angle and the anomalous reflection only appears for $\pm 2^\circ$ around the normal incidence, while the surface  behaves as a mirror for other incidence angles. However, in the conventional design the power is always coupled to $n=\pm 1$ modes and the amount of power reflected in the specular direction is small. The maximum efficiency of this design is achieved at $\theta_{\rm i}=\pm 28^\circ$. The reflection angle  can be calculated as $\theta_{\rm r}=\arcsin(\sin\theta_{\rm i}+n\sin 70^\circ)=\mp 28^\circ$.  This particular case corresponds to the retro-reflection scenario where all the energy is sent back in the  direction of the incident plane wave.  It is also worth noting that retro-reflection is the only scenario where the interaction between the two existing plane waves does not produce power modulation and the conventional linear phase gradient produces perfect results.

Finally we note that in any actual realization of inhomogeneous reactive metasurfaces the surface  elements are strongly coupled through reactive near fields, so that the local design of inhomogeneous surface reactance profile is in practice hardly possible,
although the target surface impedance is local.

\subsection{Non-local design, physical implementation and experimental validations}
\label{sec:non_local}

The above  results show the capability of surface waves propagating along an engineered gradient-phase metasurface to redirect energy and emulate the ideal ``active-passive'' behaviour needed for the implementation of perfect reflectarrays.
These results are based on the assumption that the metasurface behaves as a local impedance boundary  where no fields are allowed behind this boundary. Within  this model, we can modify punctually one element of the metasurface without affecting the characteristics of the neighbors.  However, this idealistic model does nor provide guidelines for practical designs, because  actual realizations of reactive surfaces require the use of some physical structures having a finite thickness, so that there are fields behind the mathematical metasurface boundary which  couple the elements. In this case the constituent elements cannot be designed individually using the model of periodical arrays of each element.

\begin{figure*}
	\centering
	\begin{minipage}{0.3\linewidth}
		\raisebox{1.0cm}{\subfigure[]{\includegraphics[width = 1\linewidth, keepaspectratio=true]{DiazRubioFIGr4_A.pdf}\label{fig:DiazRubioFIGr4c_patches}
		}}
		\raisebox{1.0cm}{\subfigure[]{\includegraphics[width = 1\linewidth, keepaspectratio=true]{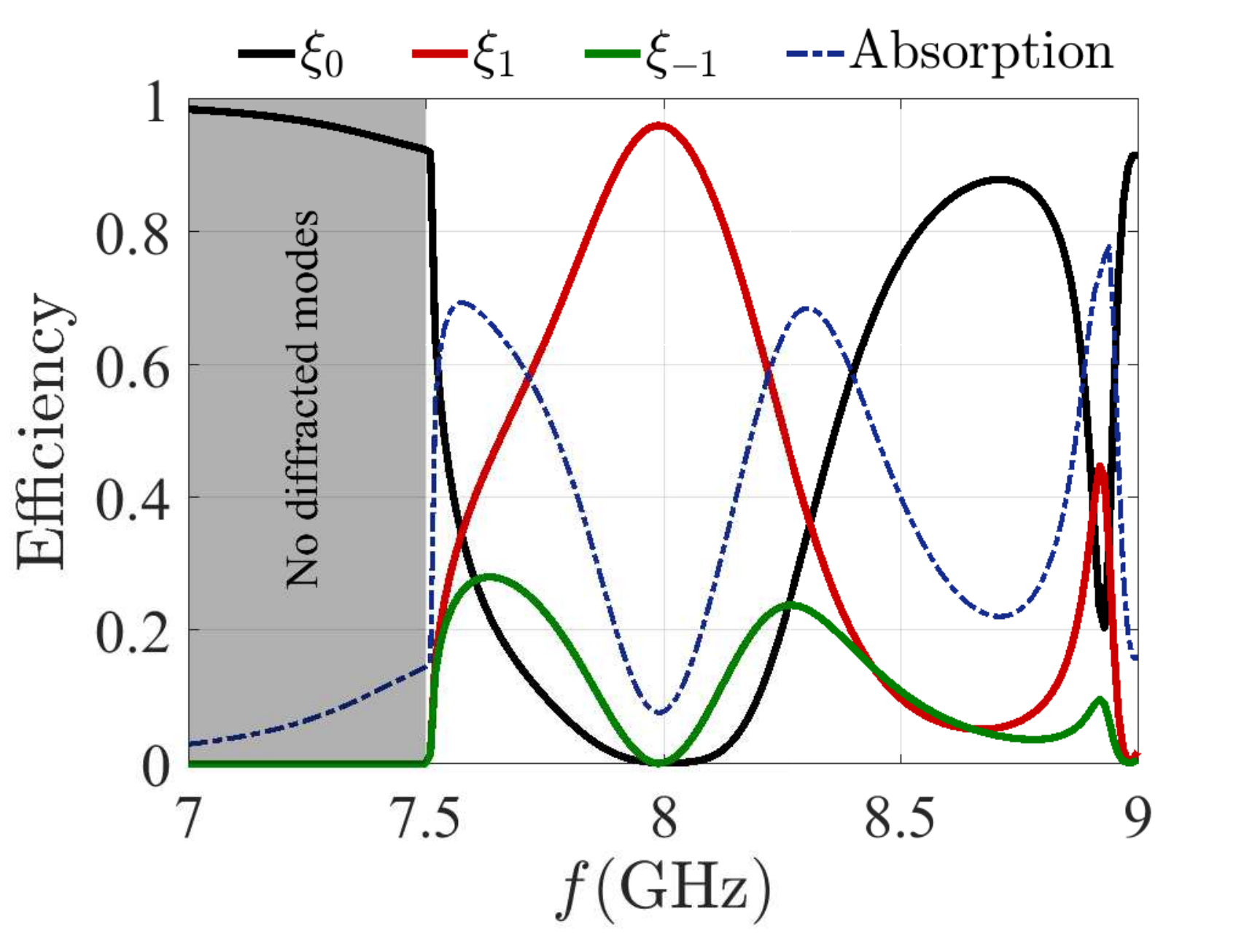}\label{fig:DiazRubioFIG4c_bandwidth}
		}}
	\end{minipage} 
	\begin{minipage}{0.3\linewidth}
		\subfigure[]{\includegraphics[width = 1\linewidth, keepaspectratio=true]{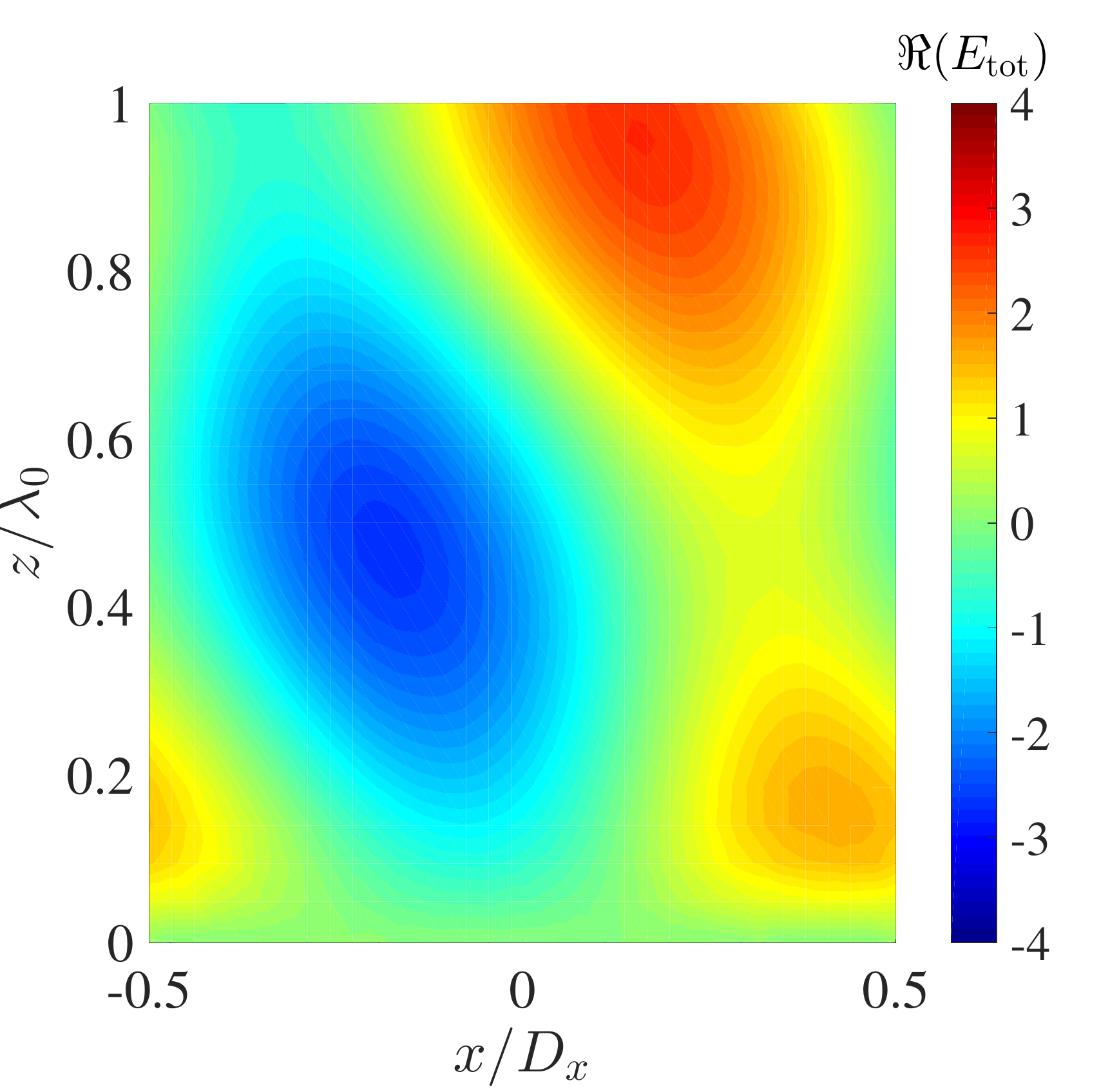}\label{fig:DiazRubioFIG4a}}
		\subfigure[]{\includegraphics[width = 1\linewidth, keepaspectratio=true]{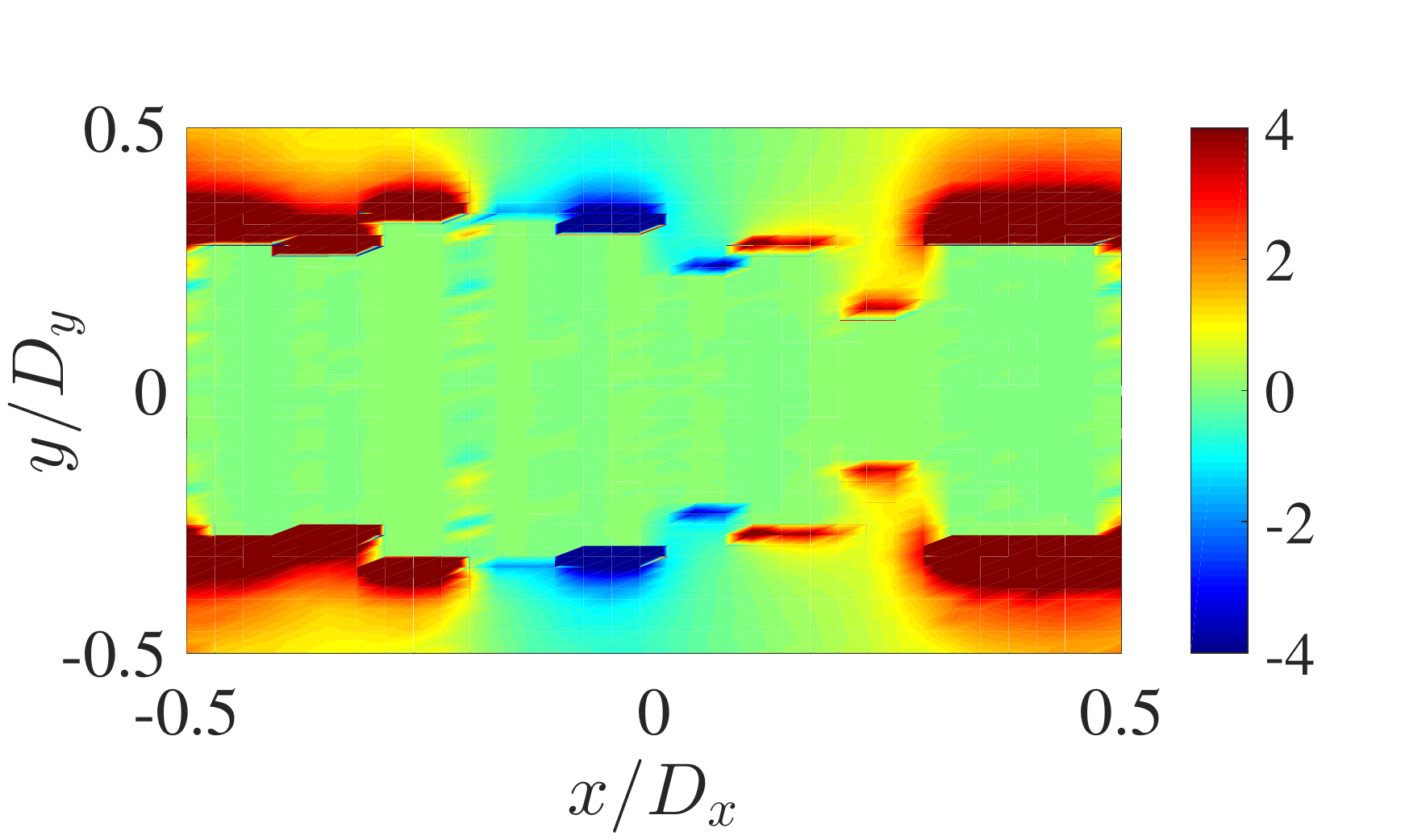}\label{fig:DiazRubioFIG4b}}
		
	\end{minipage}
	\begin{minipage}{0.3\linewidth}
		\subfigure[]{\includegraphics[width = 1\linewidth, keepaspectratio=true]{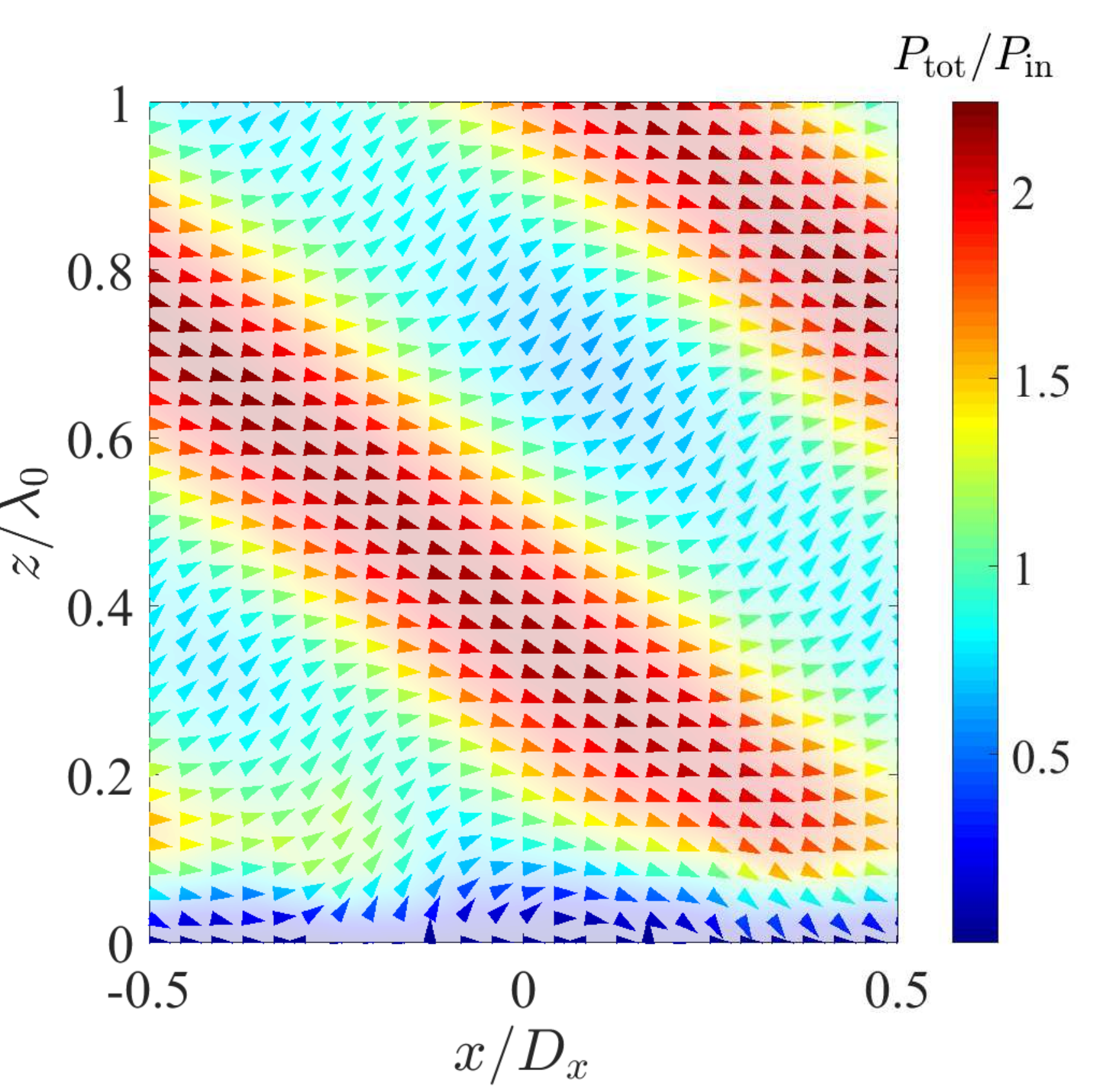}\label{fig:DiazRubioFIG4d}}
		\subfigure[]{\includegraphics[width = 1\linewidth, keepaspectratio=true]{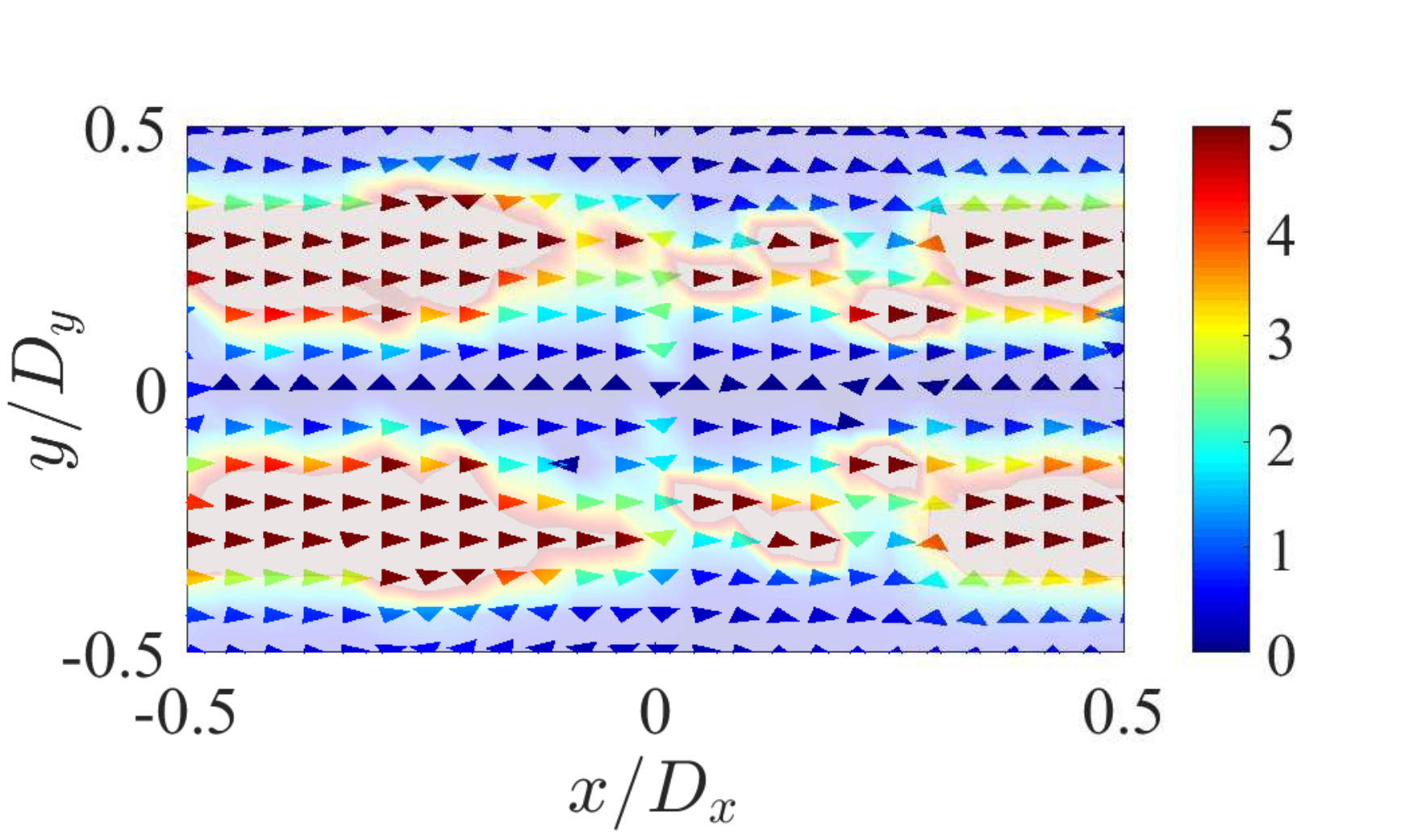}\label{fig:DiazRubioFIG4e}}
	\end{minipage} 

	\caption{(a) Proposed inhomogenerous non-local leaky-wave reflector. (b) Frequency bandwidth of the proposed design.  Simulated real part of the total electric field (c) on the $xz$-plane when $y=0$  and (d) on the $xy$-plane when $z=0$.  Real part of the total Poynting vector on (e) the $xz$-plane when $y=D_y/2$  and real part of the Poynting vector (f) on the $xy$-plane when $z=0$.  }\label{fig:DiazRubioFIG5}
\end{figure*}

We approach  the problem of realization of the required inhomogeneous leaky-wave surface  considering  one of the simplest reactive impedance surfaces: a subwavelength array of metallic patches above a metal ground plane. 
A schematic representation of the proposed system is shown in Fig.~\ref{fig:DiazRubioFIGr4c}. The modulation of the field is done by changing the length of the patches. In order to obtain the proper response of the whole unit cell, we use a local estimation according to the phase gradient dictated by Eq.~\ref{eq:active} as an initial guess (not considering the magnitude of the reflection coefficient) and carry out an optimization process for engineering the   interactions between the elements and ensuring the desired non-local response of the surface. It is important to note that the optimization process does not aim to reproduce the local response illustrated in Fig.~\ref{fig:DiazRubioFIG3}, while we aim to design the array of elements which will produce the overall ``active-lossy'' behavior described by Eq.~\ref{eq:active}. More information about the design process is available in section Methods.

Particularly, our design contains 10 metal patches per unit cell, all of them with the same width and centred along the $y=0$ line, \textcolor{blue}{see Fig.~\ref{fig:DiazRubioFIGr4c_patches}}. For the operational frequency of 8~GHz,  the lengths of the patches are 10.7, 	10.3, 12.3, 12, 11.8, 8.7, 10.2, 5.4, 11, and 10.9~mm. The patches are placed over a grounded dielectric slab with the permittivity  $\epsilon_{\rm r}=2.2$ and the loss tangent $\tan \delta=0.0009$. 
Figures~\ref{fig:DiazRubioFIG4a} and \ref{fig:DiazRubioFIG4d} show numerical simulations of the electric field distribution of the designed metasurface. 
Figure~\ref{fig:DiazRubioFIG4b} shows the simulated total Poynting vector distributions in the $xz$-plane when $y=D_y/2$. The Poynting vector is distributed according to our predictions, the power is guided below the patches and then it is launched into the desired direction.  The power distribution in the plane which contains the patches is shown in Fig.~\ref{fig:DiazRubioFIG4e}. We can see how the power is guided along the edges of the patches.   An analysis of the evanescent fields which allow the energy channeling is available in \cite{suppl}.   Figure \ref{fig:DiazRubioFIG4c_bandwidth} shows the simulated frequency response. The behavior is similar to that of the local design (Fig.~\ref{fig:DiazRubioFIGr4b}), although the anomalous-reflection frequency band is wider.

\begin{figure}[h!]
	\centering
		\begin{minipage}{0.35\linewidth}
	    \subfigure[]{\includegraphics[width = 0.22\textheight, keepaspectratio=true]{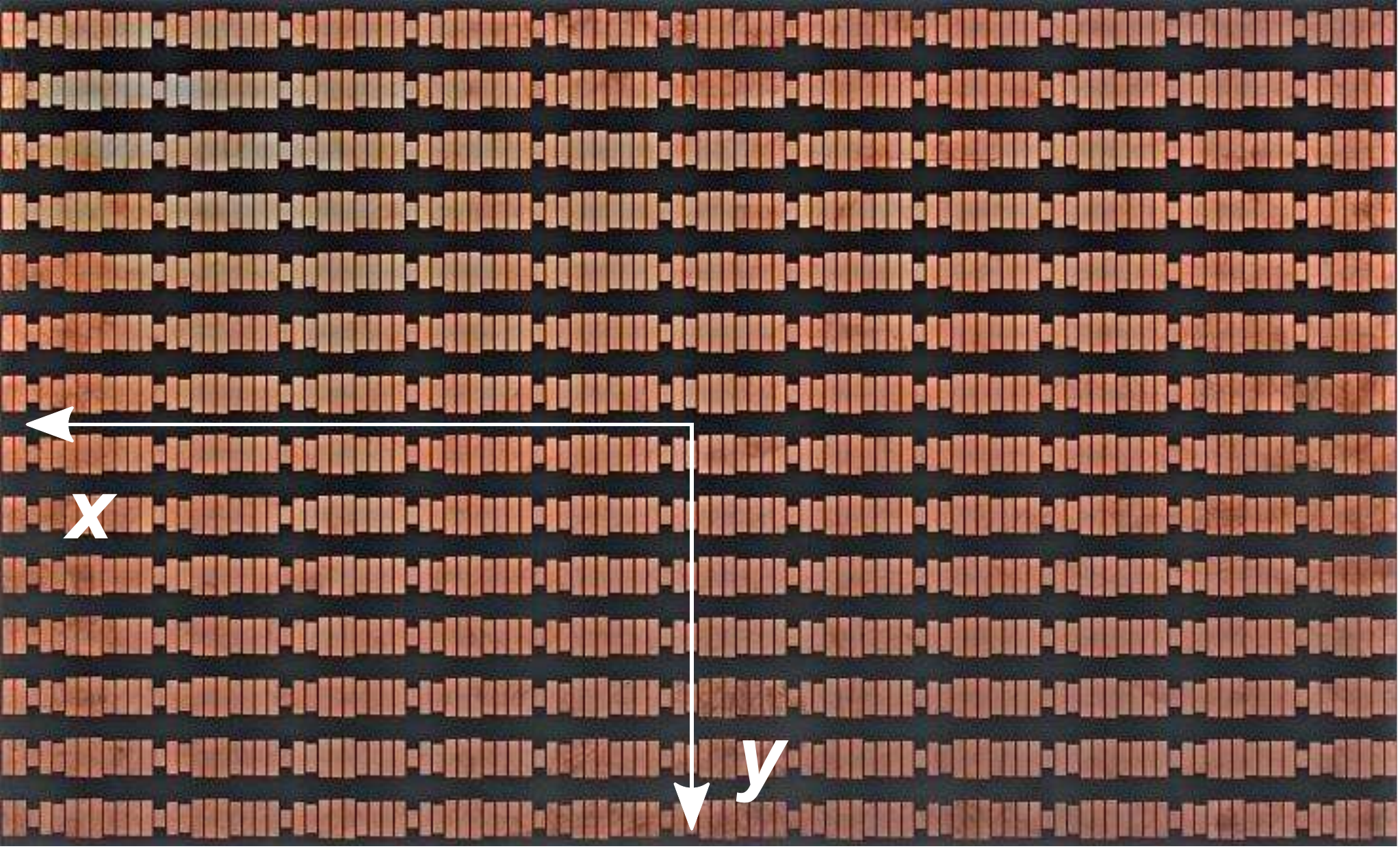}\label{fig:DiazRubioFIG4c}}
		\subfigure[]{\includegraphics[width = 0.22\textheight, keepaspectratio=true]{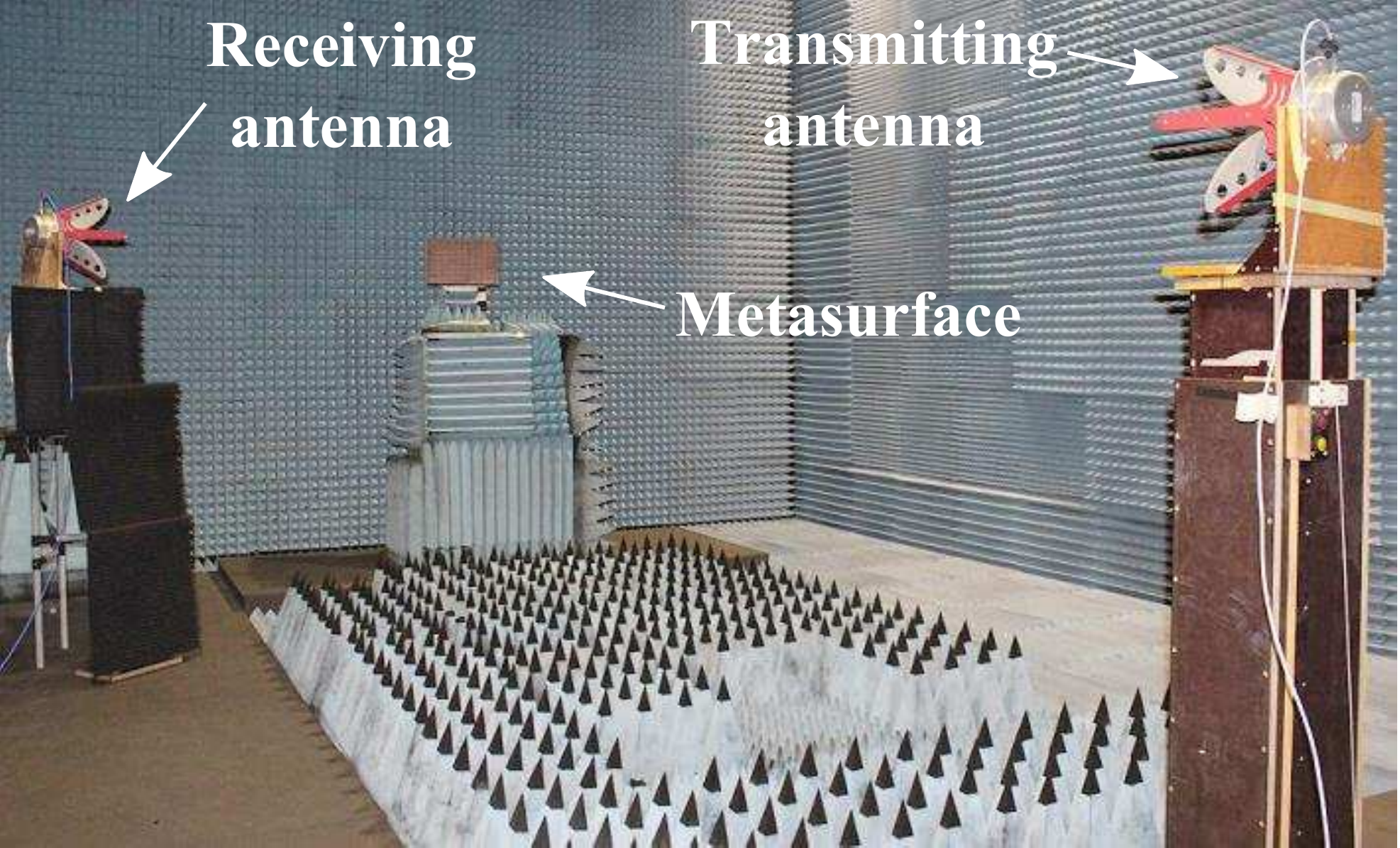}\label{fig:DiazRubioFIG4f}}
	\end{minipage}
	\begin{minipage}{0.6\linewidth}
		\subfigure[]{
		\epsfig{file=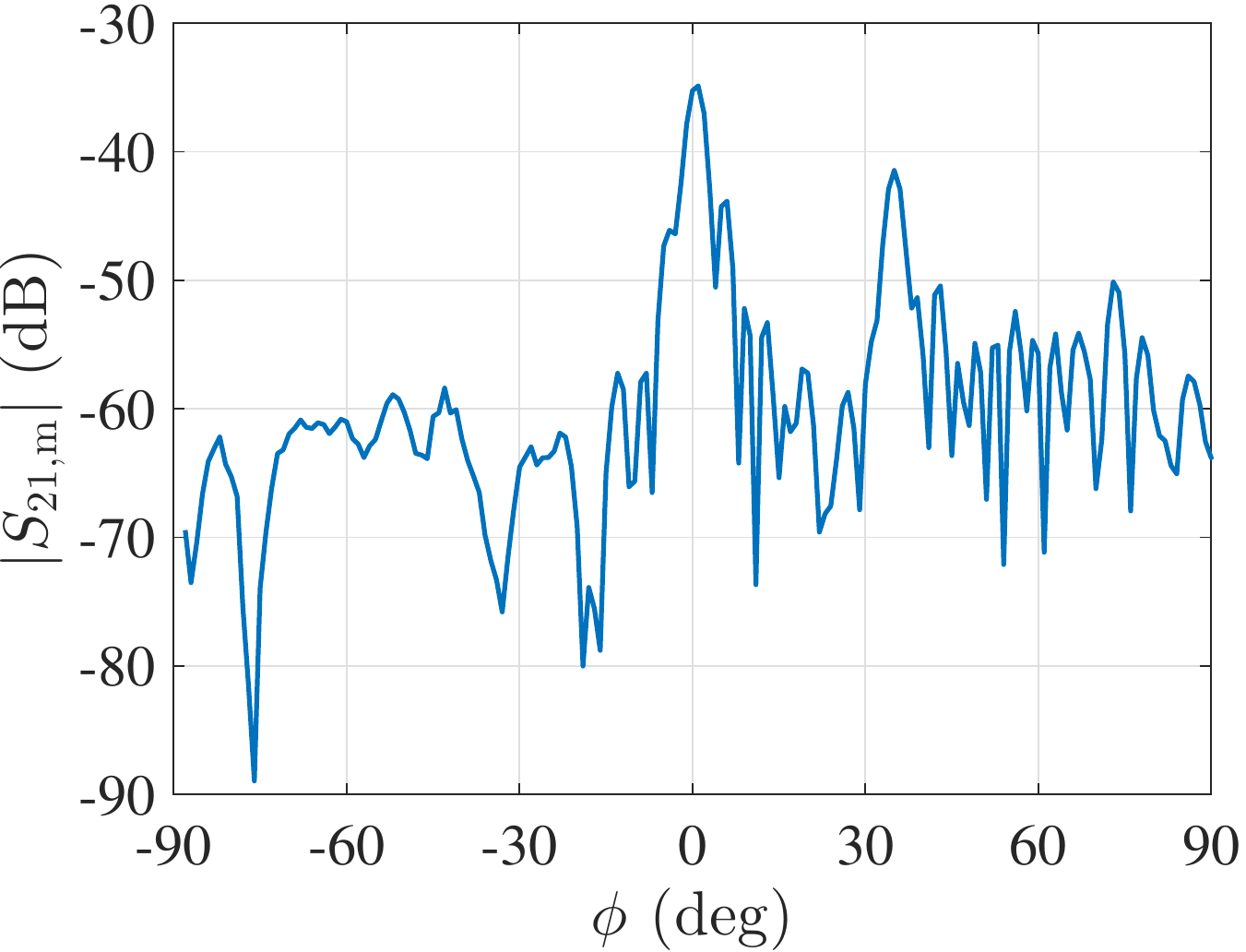, width=0.45\columnwidth}  
		\label{fig:DiazRubioFIG5a} }
	\subfigure[]{
		\epsfig{file=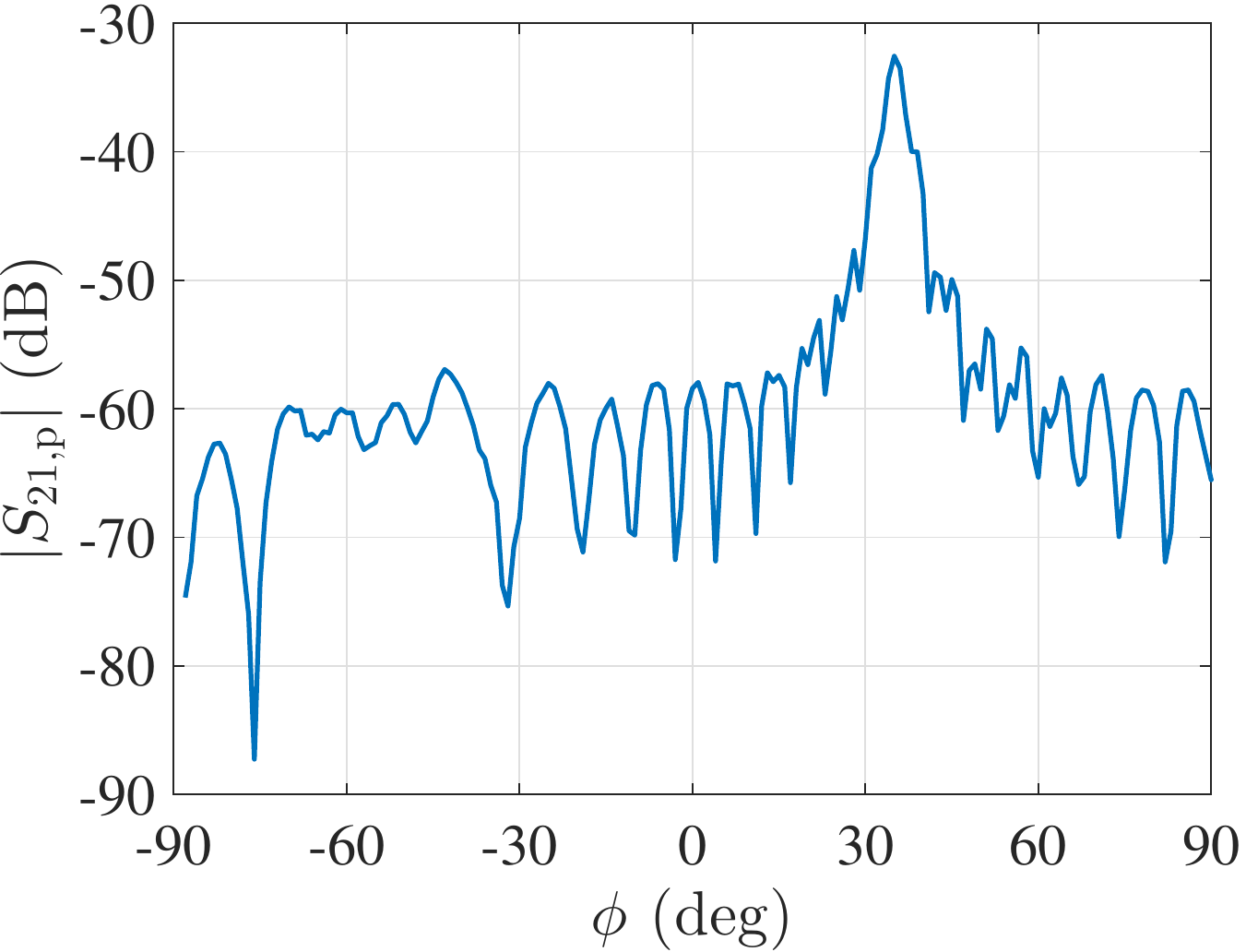, width=0.45\columnwidth} 
		\label{fig:DiazRubioFIG5b} } \\
	\subfigure[]{
		\epsfig{file=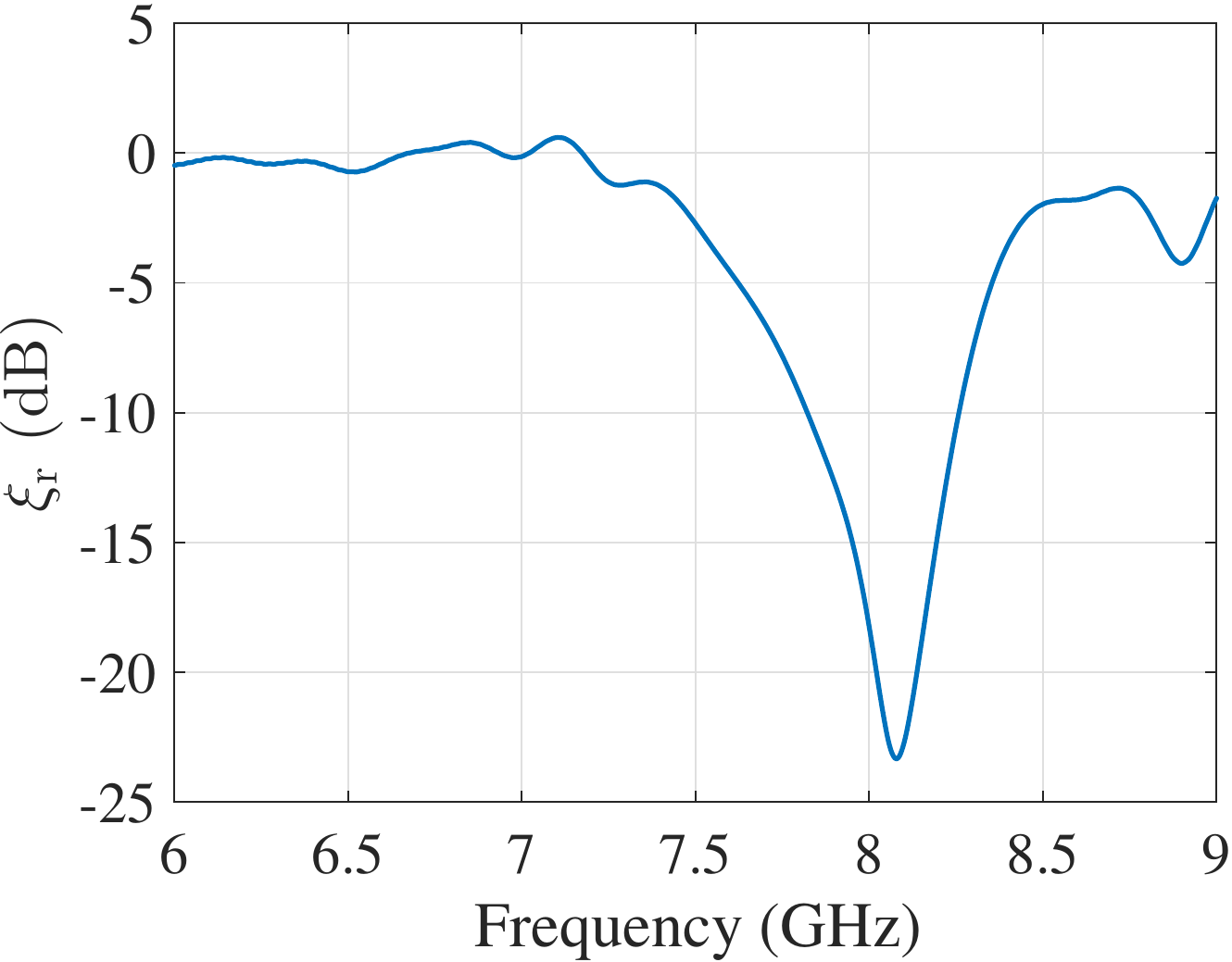, width=0.45\columnwidth}  
		\label{fig:DiazRubioFIG5c} }
	\end{minipage} 

	\caption{(a) Fabricated metasurface. (b) Experimental set-up in an anechoic chamber. Signals measured  by the receiving antenna  for different orientation angles $\phi$ of (c) the metasurface and (d) the metal plate of equal size. (e) Amplitude efficiency of the metasurface illuminated normally. 
	}\label{fig:DiazRubioFIG6}
\end{figure}

To experimentally verify the theory, we have designed and fabricated a reflectarray at 8~GHz (see Methods).
Three different experiments have been done for the experimental validation. In the first experiment, the platform with the metasurface was rotating at an angle $\phi$, while the positions of the antennas were fixed. 
Angle $\phi$ was counted from the line connecting the metasurface center and the transmitting antenna  towards the normal to the metasurface plane (positive $\phi$ corresponded to the clockwise rotation of the platform when seen from the top). 
The signal reflected from the metasurface and measured by the receiving antenna $|S_{21, {\rm m}}|$ for different angles $\phi$ is shown in Fig.~\ref{fig:DiazRubioFIG5a}. The experimental data was measured at the resonance of the metasurface occurring at 8.08~GHz.
 
The main peak of reflection towards the receiving antenna occurs when the metasurface is illuminated normally. This is an expected result meaning that in this case most of the power impinging on the surface is reflected at $70^{\circ}$ from the normal. The second peak occurring when $\phi=35^{\circ}$ corresponds to the specular reflection (incidence and reflection angles are equal) from the metamirror. This small non-zero specular reflections are acceptable  because the metamirror was designed to have zero  specular reflections only when illuminated normally. 

In order to estimate the amplitude efficiency of the metasurface $\xi_{\rm r}$ {($\xi_{\rm P}=\xi_{\rm r}^2$)}, in the second experiment we replaced the metasurface by an aluminium plate of the same size. The corresponding signal reflected from the plate and measured by the receiving antenna $|S_{21, {\rm p}}|$ versus angle $\phi$ is shown in Fig.~\ref{fig:DiazRubioFIG5b}. Now the single peak of reflection occurs when the plate is illuminated at 35$^{\circ}$, which corresponds to the specular reflection. To find the reflection efficiency of the metasurface, we normalize its signal amplitude $|S_{21, {\rm m}}|$ for $\phi=0^{\circ}$   by the signal amplitude from the reference uniform metal mirror $|S_{21, {\rm p}}|$ for $\phi=35^{\circ}$. We additionally divide the obtained value  by the correction factor $\xi_0=|S_{21, {\rm 0m}}|/|S_{21, {\rm 0p}}|$ which gives the ratio between the theoretically calculated signal amplitudes from an \textit{ideal}  metasurface (of the same size and made of lossless materials) and  a perfect conductor plate. The correction factor $\xi_0$ is less than unity because in this scenario the radiating effective area of the perfect conductor plate is greater than that of the ideal reflecting metasurface due to a different orientation with respect to the antennas. At the frequency 8.08~GHz in our particular configuration, the correction factor is equal to $\xi_0=-2.41$~dB \cite{suppl}. Thus, the reflection efficiency of the metasurface is calculated as
\e \xi_{\rm r}=\frac{1}{\xi_0}\frac{|S_{21, {\rm m}}(\phi=0^{\circ})|}{|S_{21, {\rm p}}(\phi=35^{\circ})|}=-0.28~{\rm dB}.\f
In the linear scale and expressed in terms of power, the reflection efficiency reaches 93.8\%. This result is in excellent agreement with the efficiency 94\% obtained using numerical solver (see Table~\ref{tab:Comparison}). The remaining 6.2\% of  power incident on the metasurface is mainly absorbed by it. 

In the third measurement, we fixed the orientation of the metasurface   at $\phi=0^{\circ}$  and utilized the transmitting antenna (fixed at the same position as in previous measurements) as both the transmitter and receiver. Using conventional time gating post-processing procedure, we could filter out all parasitic reflections received by the antenna (from the walls, absorbers, and due to the impedance mismatch between the antenna and the cables), retaining only the  signal reflected from the metasurface.  The measured signal amplitude was normalized by the corresponding amplitude from the reference uniform aluminium plate which was measured likewise. This ratio,  plotted in Fig.~\ref{fig:DiazRubioFIG5c}, represents the level of specular reflection from the metasurface $\xi_{\rm r}$ when illuminated normally. As is seen, at the resonance frequency 8.08~GHz the specular reflection is $\xi_{\rm r}=-23.33$~dB, which corresponds to 0.5\% of the  incident power. This result additionally confirms that the normally illuminated metasurface reflects all the power at the desired angle $70^{\circ}$. As is seen from Fig.~\ref{fig:DiazRubioFIG5c}, at frequencies below 7.2~GHz where the metal strips on the substrate are weakly excited, the metasurface behaves as a usual mirror obeying the simple reflection law.

\section{Discussion}
We have demonstrated that by proper design of planar inhomogeneous low-loss reflectors it is possible to realize conceptually perfect anomalous reflection, transforming a plane wave coming from an arbitrary direction into a single plane wave propagating into any other direction. The introduced approach fully removes the fundamental limitations on the performance of known reflectarray antennas and known metasurfaces designed with the use of the generalized Snell's law. The ``active-lossy'' behavior of conceptually perfect anomalously reflecting metasurfaces, caused by  power oscillations associated with the coexistence of two interfering plane waves in the same media,   was realized with the use of carefully engineered effects of strong spatial dispersion in an inhomogeneous leaky-wave structure.
Although in this first demonstration we assumed that the incidence is a single plane wave, the proposed physical mechanism of non-local reflections allows generalizations for arbitrary illuminations. Indeed, any arbitrary source and any metasurface response (desired reflected field) can be expressed in terms of plane wave expansions  which will correspond to power modulations and require an appropriate non-local  response \cite{arbitrary1}. 
For this reason, this work opens possibilities for creation of various metasurfaces for shaping waves, such as holograms, focusing metasurfaces or thin-sheet antennas without loosing power for parasitic scattering. 
In addition, the simple topology of metal patches printed on a thin dielectric substrate makes the proposed designs attractive in practical applications. 

\section{Methods}
\subsection{Modelling of reflective metasurfaces based on the surface impedance}
To verify the behaviour of different reflective metasurfaces we used numerical simulations in ANSYS Electromagnetic Suite 15.0.2 (HFSS 2014.0.2). 
The simulation domain is $D_x\times\lambda/10\times 2\lambda$  (along the $x$, $y$, and $z$ directions, respectively), and it corresponds to one period of the metasurface contained within master and slave boundaries in the $x$ and $y$ directions.
The surface impedance is modelled as a piece-wise constant using a discrete number of elements. Each element implements the impedance boundary dictated by Eqs.~(\ref{eq:conventional}),  (\ref{eq:lossy}), or (\ref{eq:active}). The number of elements per period is 50 for the conventional design, 18 for the lossy, and 8 for the active design. The effect of the discretization was studied and the number of elements was chosen in order to ensure the accuracy of the results.  
The system is illuminated by a TE polarized plane wave.  Two different simulations have been done: plane-wave excitation (extraction of the scattered fields, results represented in Fig.~\ref{fig:DiazRubioFIG2}) and Floquet port excitation (study of the power scattered into each Floquet mode, results shown in Table~\ref{tab:Comparison}).

For the inhomogeneous leaky-wave antenna surface (results shown in Fig.~\ref{fig:DiazRubioFIG3}), the number of elements is 15. For the optimization process, we used the optimization tool of HFSS and the Quasi Newton (Gradient) algorithm with the goal condition $S_{11}=1$. 

\subsection{Design and modelling of a perfect reflectarray}

The prototype presented in this work has been designed for operation at 8~GHz. 
The unit cell consists of  10   rectangular copper patches above a copper ground plane. The width of all the patches is $w=3.5$~mm. The thickness of the dielectric substrate is $1.575$~mm,  and final dimensions of each unit cell in the $xy$-plane are  $D_x=40$~mm and $D_y=18.75$~mm.

The first step in the design methodology was to determine the required reflected field for each element. To do that, we use the phase gradient defined by Eq.~(\ref{eq:active}). Then we periodically arrange each element (using a homogeneous array model) and calculate the length that produces the desired phase shift. The simulation domain was $D_x/10\times D_y\times 2\lambda$  (along the $x$, $y$, and $z$ directions). The dielectric material used in the simulations was  Rogers~5880, with  $\epsilon_{\rm r}=2.2$, $\tan \delta=0.0009$, and the thickness  $1.575$~mm. The rectangular patches and the ground plane were modelled as cooper ($\sigma=58 \times 10^6~\rm{ S/m}$) with the thickness $70$~$\rm{\mu m}$.

Once we knew the dimensions of all the elements in the unit cells, the second step was the optimization of the complete unit cell which consists of 10 different patches. We did a numerical optimization of the structure that corrects the effects produced in the non-homogeneous array that were not accounted for in the initial locally homogeneous approximation. The simulation domain of the complete unit cell was $D_x\times D_y\times 2\lambda$. The response of the inhomogeneous leaky-wave surface with 10 elements has been optimized with HFSS. The Quasi Newton (Gradient) algorithm was used and the goal was defined as $S_{11}=1$. The optimized  dimensions of the patches are given in Section~\ref{sec:non_local}. 

\subsection{Perfect reflectarray realization and measurement}

The metasurface sample designed to operate at the frequency of 8~GHz was manufactured using conventional printed circuit board technology on a $1.575$~mm thick Rogers~5880  substrate. The sample comprises 11 unit cell along the $x$-axis and 14 unit cells along the $y$-axis [see Fig.~\ref{fig:DiazRubioFIG4c}] and has the size of $11.7\lambda=440$~mm and $7\lambda=262.5$~mm, respectively. 

The operation of the designed non-local reflecting metasurface was verified by  measurements in an anechoic chamber emulating the free-space environment. A vector network analyzer was connected to a transmitting quad-ridged horn antenna with 11~dBi gain at 8~GHz [see Fig.~\ref{fig:DiazRubioFIG4f}]. The metasurface was located at a distance of 5.5~m (about $147\lambda$)  from the transmitting antenna where the radiation from the antenna can be approximated as a plane wave. To control the metasurface orientation, it was attached to a  platform rotating around the $y$-axis. 
The receiving antenna, identical to the transmitting one, was positioned at a distance 2.387~m (about $64\lambda$)  from the center of the metasurface. Both  antennas and the metasurface form in space  a triangle with the angle $70^{\circ}$ at the metasurface center. The measured results are presented in Section~\ref{sec:non_local}.

\section*{Acknowledgment}
This work was supported in part by the Academy of Finland (project 287894).
The authors would like to thank Muhammad Ali and Abbas Manavi for technical help with the experimental equipment.

 All data needed to evaluate the conclusions in the paper are present in the paper and/or the Supplementary Materials. Additional data related to this paper may be requested from the authors.

\section*{Author contributions}
A. DR. performed the numerical calculations. A.DR. and V.A. designed the samples. V.A and A.E conducted the experiment and analyzed the measurements. A. DR., V.A., and S.T. wrote the paper. S.T. supervised the project. All authors contributed to the scientific discussion of the manuscript.

\section*{Competing interests} The authors declare that they have no competing interests.

\end{document}